\documentclass[twocolumn,pra,aps,superscriptaddress, showpacs, amsmath,amssymb,10pt]{revtex4-1} 
\usepackage{amsmath} 
\usepackage{amssymb} 
\usepackage{amsfonts} 
\usepackage{bm} 
\usepackage{amssymb} 
\usepackage{graphicx} 
\usepackage{dcolumn} 
\usepackage{txfonts} 
\usepackage{wasysym} 
\usepackage{makeidx}
\usepackage{color}
\usepackage{mathtools}
\usepackage{threeparttable}
\usepackage[linkcolor=blue,anchorcolor=black,citecolor=blue,colorlinks=true]{hyperref}
\usepackage{breakurl}

\begin{document}
\title{Stability Improvement of Nuclear Magnetic Resonance Gyroscope with Self-Calibrating Parametric Magnetometer}
\author{Guoping Gao}
\affiliation{Beijing Computational Science Research Center, Beijing 100193, PR China}%
\author{Jinbo Hu}
\affiliation{Beijing Computational Science Research Center, Beijing 100193, PR China}%
\author{Feng Tang}
\affiliation{Beijing Computational Science Research Center, Beijing 100193, PR China}%
\author{Wenhui Liu}
\affiliation{Beijing Computational Science Research Center, Beijing 100193, PR China}%
\author{Xiangdong Zhang}
\affiliation{Beijing Computational Science Research Center, Beijing 100193, PR China}%
\author{Baoxu Wang}
\affiliation{Institute of Systems Engineering, China Academy of Engineering Physics, Mianyang 621999, Sichuan, PR China}%
\author{Dongge Deng}
\affiliation{Institute of Systems Engineering, China Academy of Engineering Physics, Mianyang 621999, Sichuan, PR China}%
\author{Mingzhi Zhu}
\affiliation{Institute of Systems Engineering, China Academy of Engineering Physics, Mianyang 621999, Sichuan, PR China}%
\author{Nan Zhao}
\email{nzhao@csrc.ac.cn}
\affiliation{Beijing Computational Science Research Center, Beijing 100193, PR China}%
\date{\today}

\begin{abstract}
In this paper, we study the stability of nuclear magnetic resonance gyroscope (NMRG), which employs Xe nuclear spins to measure inertial rotation rate.
The Xe spin polarization is sensed by an in-situ Rb-magnetometer. 
The Rb-magnetometer works in a parametric oscillation mode (henceforth referred to as the Rb parametric magnetometer, or Rb-PM), 
in which the Larmor frequency of the Rb spins is modulated and the transverse components of Xe nuclear spin polarization are measured.
As the measurement output of the Rb-PM, the phase of the Xe nuclear spin precession is eventually converted to the Xe nuclear magnetic resonance (NMR) frequencies and the inertial rotation rate. 
Here we provide a comprehensive study of the NMR phase measured by the Rb-PM, and analyze the influence of various control parameters, 
including the DC magnetic field, the frequency and phase of the modulation field, and the Rb resonance linewidth, on the stability of the NMR phase.
Based on these analysis, we propose and implement a self-calibrating method to compensate the NMR phase drift during the Rb-PM measurement.
With the self-calibrating Rb-PM, we demonstrate a significant improvement of the bias stability of NMRG.
\end{abstract}


\maketitle
\section{Introduction}
Nuclear magnetic resonance gyroscope (NMRG) was proposed in 1970s \cite{Grover1979} after the discovery of the spin-exchange optical pumping (SEOP) of nuclear spins \cite{PhysRevLett.5.373}.
Great interest in NMRG revived in recent years \cite{6243606,Meyer2014,Walker2016}, because of the need of inertial measurement devices with high precision and high portability.
Although a compact prototype of NMRG system based on dual-species Xe nuclear spins was successfully demonstrated in 2010s \cite{Walker2016}, 
great efforts were made in the past a few years to further improve the performance (e.g., sensitivity and stability) of NMRGs \cite{korverSynchronousSpinExchangeOptical2015,limesHe3Xe1292018,haoHerriottcavityassistedClosedloopXe2021}.

A typical NMRG system consists of two types of spins, namely, the nuclear spins of noble gas atoms and the atomic spins of alkali-metal vapor (Xe atoms and Rb atoms in this paper).
The Xe nuclear spins are used to discriminate the inertial rotation rate utilizing their long coherence time.
The Rb atomic spins create nuclear spin polarization of Xe atoms via the SEOP process, and serve as an in-situ magnetometer, converting the nuclear spins precession information to voltage signal.

When confined in a mm-sized glass cell, the mixed ensemble of Rb-Xe spins is a good candidate for a portable inertial sensing device.
The polarized Xe nuclear spins precess about the magnetic field with a stable frequency.
The long spin coherence time $T_2\sim 10~{\rm s}$ corresponds to a resonance frequency uncertainty of $\Delta \nu = 1/(2\pi T_2)\sim 10^1~{\rm mHz}$.
The Xe nuclear spins under resonant driving create an oscillating magnetic field with amplitude $B_{\rm Xe}\sim 10~{\rm nT}$ to the Rb atomic spins.
The Rb atomic spins, when modulated properly, work as a magnetometer with a typical sensitivity $\lesssim 1~{\rm pT}/\sqrt{\rm Hz}$.
This sensitivity gives rise to a high signal-to-noise ratio (${\rm SNR}\sim 10^4~\sqrt{\rm Hz}$) of the Xe field $B_{\rm Xe}$ measurement.
The narrow line width $\Delta \nu$, together with the the high SNR, allows the NMRG to measure the inertial rotation rate with an uncertainty as low as $\sim  0.1~{\rm  \mu Hz}$ within an averaging time of $\sim 10^2~{\rm s}$.

The sub-${\rm \mu Hz}$ high-precision inertial rate measurement requires the high stability of the NMRG system.
Inevitable noise from environment or imperfect control of the NMRG system disturbs the dynamics of both Xe and Rb spins. 
The disturbance, usually in low-frequency range, causes the long-term drift of the measurement results and, eventually, limits the precision of the NMRG.
In general, the Xe nuclear spins and the Rb atomic spins suffer from disturbance of different physical origins.

The nuclear spins of Xe are primarily influenced by the drift of the magnetic fields. 
The dual-species NMRG employs two isotopes of Xe, namely $^{129}\mathrm{Xe}$ and $^{131}\mathrm{Xe}$, to eliminate the common-mode magnetic field (such as the field generated by coils).
The differential-mode magnetic field originates from the Rb-Xe spin exchange collision \cite{PhysRevLett.111.102001,8854116}. 
The Rb atomic spins create effective magnetic fields to the Xe nuclear spins, which are referred to as the Rb {\it polarization fields} hereafter, thus changing the nuclear spin precession frequency of the two isotopes.
The effective Rb polarization fields felt by $^{129}{\rm Xe}$ and $^{131}{\rm Xe}$ isotopes are slightly different, typically, by an amount of $\sim 10^1~{\rm pT}$. 
The differential polarization Rb field is varying if the NMRG system is not well-controlled (e.g., if the cell temperature drifts). 
The uncontrolled change of the differential polarization field (typically in the order of $\sim {\rm pT}$ or lower) is regarded as one of the main reason limiting the stability of the NMRG.
Great efforts have been made to develop NMRG systems which are immune to the drift of the differential polarization field, 
including nulling the polarization field by periodically flipping the Rb spins \cite{korverSynchronousSpinExchangeOptical2015,limesHe3Xe1292018,PhysRevLett.111.043002},
and cancelling the polarization field by introducing extra degrees of freedom \cite{zhangStableAtomicMagnetometer2023a}.

Besides the Xe nuclear spins, the Rb magnetometer also contributes to NMRG instability.
Indeed, the stability of the Rb magnetometer and that of the entire NMRG system are inextricably linked. 
Previous studies have primarily focused on the short-term sensitivity of the Rb magnetometer  \cite{Budker2007,seltzer2008}, 
whereas a systematic investigation into the low-frequency behaviors of in-situ Rb magnetometers within NMRG systems is currently lacking.
In this paper, we present a comprehensive study of the stability of the Rb magnetometer of the NMRG. 
Based on the transfer function method, we give a theoretical framework of the noise analysis of the NMRG system, 
with a special attention on the low-frequency phase noise introduced by the Rb atomic spins under parametric modulations (i.e., the Rb parametric magnetometer, or, the Rb-PM).
We further investigate the physical origin of phase noise induced by the Rb-PM with the exact solution of the equation of motion governing Rb atomic spins. 
Analytic expressions for the dependence of NMR phase measured by Rb-PM on various control parameters are obtained and experimentally verified.
These parameters includes the static magnetic field $B_0$, the AC modulation field phase $\theta_{\rm ac}$ and the Rb spin relaxation rate $\Gamma_{\rm Rb}$. 
Based on these analysis, we propose and demonstrate a self-calibrating approach to mitigate the phase drift caused by Rb-PM. 
Our results indicate that the long-term stability of NMRG is significantly improved with this self-calibrating Rb-PM method.

This paper is organized as follows. The experimental setup and the characteristic parameters of the NMRG system are introduced in Section~\ref{Sect:NMRG_setup}.
Section~\ref{Sect:Noise} presents the theoretical analysis of the noise propagation in the NMRG system.
The phase noise introduced by the Rb-PM is studied in Section~\ref{Sect:PM}.
The stability improvement of the NMRG with the self-calibrating Rb-PM is demonstrated in Section~\ref{Sect:NMRGStability}, and the conclusion and outlook is presented in Section~\ref{Sect:Conclusion}.
\section{NMRG Experimental Setup }
\label{Sect:NMRG_setup}
\subsection{NMRG setup}

\begin{figure}[tb]
    \centering
    \includegraphics[scale=1]{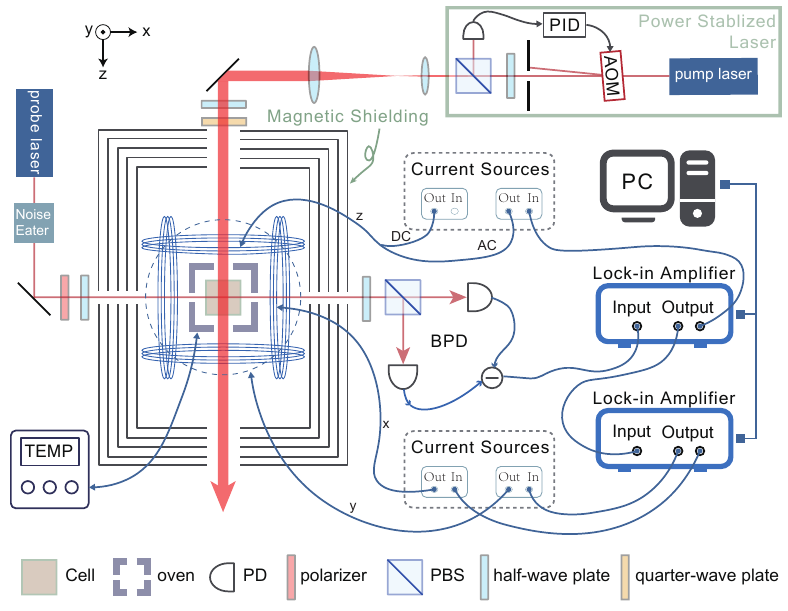}
    \caption{Schematic illustration of the experimental setup. AOM: acousto-optic modulator; PBS: polarization beam splitter; 
    PD: photo-detector; BPD: balanced photo-detector; PID: proportion-integration-differentiation controller; TEMP: temperature controller.}
    \label{fig:setup}
\end{figure}

We establish an NMRG setup as sketched in Fig.~\ref{fig:setup}. 
A cubic glass cell with inner side length $L=8~{\rm mm}$ is placed in an oven made of boron nitride.
The cell is filled with Rb atoms of natural abundance, $450~{\rm Torr}$ ${\rm N}_2$, $4~{\rm Torr}$ $^{129}{\rm Xe}$, and $14~{\rm Torr}$ $^{131}{\rm Xe}$.
The oven is heated by high-frequency ($\sim 400~{\rm kHz}$) AC current.
The magnetic field along $z$ direction is created by two sets of Helmholtz coils, one for the DC magnetic field, and the other for the AC modulation field.
A large inductance ($\sim 1~{\rm H}$) is connected in series with the DC coil to suppress the electromotive voltage from the AC coil.
The DC current is generated by an home-made low-noise current source based on the Libbrecht-Hall design \cite{Libbrecht1993}.
The transverse fields along $x$ and $y$ directions are created by two sets of saddle coils.
The oven and the coils are placed in a five-layer $\mu$-metal magnetic shield.
The Rb atomic spins are optically pumped by a $\sigma^{+}$-polarized $795~{\rm nm}$ laser beam along the $z$ direction.
The angular momentum of the polarized Rb atoms are then transferred to the Xe nuclear spins via the Rb-Xe spin-exchange collisions \cite{walker1997}.
The hyper-polarized Xe nuclear spins creates an effective field, which is detected by an in-situ Rb magnetometer \cite{liParametricModulationAtomic2006, eklundMicrogyroscopeBasedSpinPolarized2008, Walker2016}.
The transverse spin polarization of Rb atoms are detected from via the Faraday rotation effect by a linearly polarized probe beam along the $x$ direction. 
The rotation of the polarization plane of the probe beam is measured by a balanced photo detector (BPD),
and the output voltage signal is digitized and analysed by a two-stage lock-in amplification system (see the details below).

\subsection{Characteristic parameters}
The NMRG system consists of two subsystems, namely, the Rb-PM subsystem and the Xe NMR subsystem.
The Rb-PM subsystem converts the Xe spin precession to a voltage signal, whose working principle will be presented in detail in Sect~\ref{Sect:PM}.
The Xe NMR subsystem is responsible to sense the inertial rotation by the change of the resonance frequencies of the Xe nuclear spins.
Two isotopes, $^{129}{\rm Xe}$ and $^{131}{\rm Xe}$ with different gyromagnetic ratios, are used to eliminate frequency shift induced by the fluctuation of the magnetic field.
In this section, we present characteristic parameters of the Rb-PM subsystem and the Xe NMR subsystem, which are crucial to the NMRG performance.

The magnetic field sensitivity is an essential parameter of the Rb-PM subsystem.
The sensitivity is determined by the Rb magnetic resonance linewidth and the background noise level.
In our setup, with the cell temperature at $T=95^{\circ}$C and the pump beam of about $P=100~{\rm mW}$ power, the magnetic resonance linewidth of the Rb atomic spins is $\Gamma_{\rm Rb}\sim 2\pi \times 7.5~{\rm kHz}$ (half-width at half-height). 
The linewidth is mainly determined by the Rb-Xe spin-exchange collision and the optical pumping processes \cite{seltzer2008,PhysRevA.104.023105,nelsonRbXeSpinRelaxation2001,Appelt1998,Zeng1985}.
Limited by the shot-noise background of the probe beam, the magnetic field sensitivity of the Rb-PM is  $0.66~{\rm pT/\sqrt{\rm Hz}}$. 

The performance of the Xe NMR subsystem is characterized by the $T_2$ times of $^{129}{\rm Xe}$ and $^{131}{\rm Xe}$, and their nuclear spin polarization.
The $T_2$ times measured in our system are $T_{2,{\rm Xe}}^{(129)} = 5.8~{\rm s}$ and $T_{2, {\rm Xe}}^{(131)}= 11.0~{\rm s}$, for $^{129}{\rm Xe}$ and $^{131}{\rm Xe}$ spins respectively.
The $T_2$ time of $^{129}{\rm Xe}$ is limited by the gradient of the polarization field, while the $T_2$ time of $^{131}{\rm Xe}$ is mainly determined by the spin relaxation on the cell wall \cite{wuWallInteractionsSpinpolarized2021}.
The nuclear spin polarizations of the two isotopes are characterized by the effective magnetic fields felt by the Rb spins.
The effective magnetic field strength is determined by the spin-exchange pumping rate and the longitudinal spin relaxation rate of Xe spins.
The effective magnetic field strength are different from cell to cell.
The typical values of the total field strength are $\sim 200~{\rm nT}$ for $^{129}{\rm Xe}$ and $\sim 40~{\rm nT}$ for $^{131}{\rm Xe}$.
With an resonance AC driving field, the size of the transverse components in our NMRG system are $B_{\rm Xe}^{(129)}\sim 37~{\rm nT}$ and $B_{\rm Xe}^{(131)}\sim 9~{\rm nT}$. 

The parameters mentioned above are crucial to the NMRG short-term sensitivity.
Further optimization of the sensitivity is possible, e.g., increasing the $T_2$ time of nuclear spins by improving the homogeneity of the pump beam intensity distribution.
However, in this paper, we will focus on the long-term stability of the NMRG, which relies on the stability of various control parameters discussed in the following.

\subsection{Stability of control parameters}
A stable NMRG necessitates precise control of various parameters, including the cell temperature $T$, the power $P$ and frequency $\nu$ of the pump and probe laser beams, and the DC magnetic field strength $B_0$, etc.
In the following, we briefly summarize the method used to control these parameters and their stability achieved in our NMRG system.

The cell temperature is measured by a PT1000 resistance temperature detector (RTD),
and stabilized by a high-precision temperature controller.
The typical temperature fluctuation at the RTD is $\Delta T \lesssim \pm 0.002^{\circ}{\rm C}$.
Exactly speaking, the temperature at the RTD may be different from that inside the vapour cell.
It is reasonable to assume the actual cell temperature is stabilized within $\lesssim \pm 0.01^{\circ}{\rm C}$.
The slow drift of the cell temperature and its inhomogeneity could be the main factors that cause the drift of the NMRG.

The power and frequency of the laser beams are stabilized by close-loop feedback control systems.
A fractional part of the laser power is sampled by a beam splitter with fixed branching ratio.
Then, a PID-based control system is used to stabilize the sampled intensity.
The relative change of laser power incident into the vapour cell is $\Delta P/P \sim 10^{-3}$.
The laser frequency is monitored by a wavelength meter.
The difference between the real-time measured laser frequency and the target frequency is converted to a voltage signal, and feedback to the laser controller,
so that the  laser frequency is locked to the target frequency within a range of $\Delta \nu \sim 1~{\rm MHz}$.

The DC magnetic field along the $z$ direction is generated by a home-made current source according to Ref.~\cite{Libbrecht1993}.
The output current is $46.889~{\rm mA}$ with a long-term drift $\lesssim 0.3~{\rm \mu A}$ (over a monitoring duration of $100~{\rm s}$).
With the coil coefficient $\sim 400~{\rm nT/mA}$, the DC current drift corresponds to a $\lesssim 500~{\rm mHz}$ ($\lesssim 1~{\rm mHz}$) drift of the Larmor frequency of $^{85}{\rm Rb}$ atomic spins ($^{129}{\rm Xe}$ nuclear spins).

Besides the well-controlled conditions above, there are other factors that may affect the NMRG stability.
In particular, fluctuations in room temperature (typically $\pm 1^{\circ}{\rm C}$ within a day) can cause the drift of electronic devices.
This temperature drift could be complicated and system-dependent.
The purpose of this paper is to establish a quantitative connection between the drift of various physical quantities and the drift of the resultant NMRG signal. 
Also, we will propose a self-calibrating method to improve the NMRG stability.

\section{NMRG Noise: General Analysis}
\label{Sect:Noise}
Before delving into the stabilization method, we provide a theoretical description of the noise spectrum of the NMRG. 
Our quantitative model is established on the basis of transfer functions inherent to the NMRG system. 
While some aspects have been previously discussed \cite{Tang_2020,tang2019optimizations}, for reader convenience, we summarize prior findings and present a comprehensive theoretical framework in a unified symbolic system. 

\subsection{Transfer functions of NMRG}
\label{Sect:TFNMRG}
\subsubsection{NMR system}
In a static magnetic field $B_0$ along the $z$ direction, the $^{129}$Xe and $^{131}$Xe nuclear spins precess with Larmor frequencies (absolute values) 
\begin{eqnarray}
\Omega_{129} = \vert \gamma^{(129)}_{\rm Xe} \vert \left(B_0 + B_{\rm{A}}^{(129)}\right) + \Omega_{\rm{rot}}, \\
\Omega_{131} = \vert \gamma^{(131)}_{\rm Xe} \vert \left(B_0 + B_{\rm{A}}^{(131)}\right)- \Omega_{\rm{rot}},
\end{eqnarray}
where $\gamma^{(129)}_{\rm Xe} = -2\pi\times 11.86~{\rm mHz/nT}$ and $\gamma^{(131)}_{\rm Xe} = 2\pi \times 3.52~{\rm mHz/nT}$ are the gyromagnetic ratios, respectively,
$B_A^{(129)}$ and $B_A^{(131)}$ are the Rb polarization fields felt by the Xe isotopes, and $\Omega_{\rm{rot}}$ is the rotation rate to be measured.
Notice that the effective fields $B_A^{(129)}$ and $B_A^{(131)}$ are isotope-dependent, which are the main source of NMRG instability.

To drive the nuclear spins, a transverse driving field $B_y^{(\alpha)}(t) = B_1^{(\alpha)} \cos (\omega_{\alpha} t)$, 
with amplitude $B_1^{(\alpha)}$ and frequency $\omega_{\alpha}>0$, 
is applied along $y$ direction with $\alpha=129$ or $131$.
In the near-resonant regime, the phase $\varphi_{\rm Xe}^{(\alpha)}$ of the spin precession relates to the detuning $\Delta_{\alpha} = \Omega_{\alpha} - \omega_{\alpha}$ as \cite{Walker2016}
\begin{equation}
	\frac{d \varphi_{\rm Xe}^{(\alpha)}}{dt} = -\Gamma_{\alpha} \varphi_{\rm Xe}^{(\alpha)} + \Delta_{\alpha},
	\label{Eq:XeEoM}
\end{equation}
where $\Gamma_{\alpha} = 1/T^{(\alpha)}_{2, {\rm Xe}}$ is the transverse spin relaxation rate of Xe isotope $\alpha$.

\begin{figure}
	\centering
	\includegraphics[scale=1]{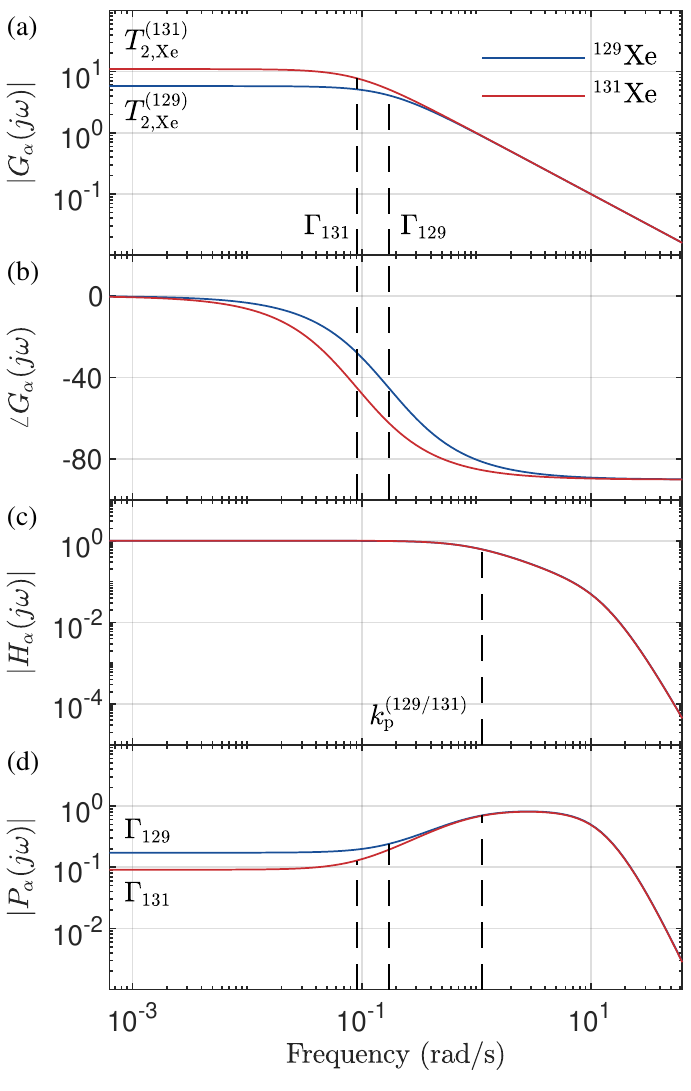}
	\caption{The frequency response of the NMRG system. (a) and (b) are the magnitude-frequency response $\vert G_\alpha(j\omega)\vert$ and phase-frequency response $\angle G_\alpha(j\omega)$ of the nuclear spins system, respectively. 
	(c) and (d) are the magnitude-frequency response of the close-loop $f$-$f$ transfer function $H_\alpha(j\omega)$ and the $\varphi$-$f$ transfer function $P_\alpha(j\omega)$, respectively.
	The parameters used in these figures are $k_{\rm P}^{(129/131)}= -1.1~ {\rm rad/s}$, $k_{\rm I}^{(129)}= -0.18~{\rm rad/s^2}$, 
	$k_{\rm I}^{(131)}= -0.087~{\rm rad/s^2}$, $T_{129} = 5.8~{\rm s}$, and $T_{131} = 11.0~{\rm s}$.}
	\label{fig:FreqResponse}
\end{figure}

The equation of motion \eqref{Eq:XeEoM} of the NMR phase implies that, 
with the input frequency detuning $\Delta_{\alpha}$ and the output spin precession phase $\varphi_{\rm Xe}^{(\alpha)}$, 
the NMR system behaves as a 1st order low-pass filter with the transfer function
\begin{equation}
	G_{\alpha}(s) = \frac{1}{\Gamma_{\alpha}+ s},
\end{equation}
where $s$ represents the complex frequency variable in the Laplace transform.
Figures~\ref{fig:FreqResponse}(a) \& \ref{fig:FreqResponse}(b) shows the magnitude-frequency and phase-frequency response of the transfer function $G_{\alpha}(s)$ for the NMR system, respectively. 

\subsubsection{Frequency-frequency transfer function}
\begin{figure}
	\centering
	\includegraphics[scale=1]{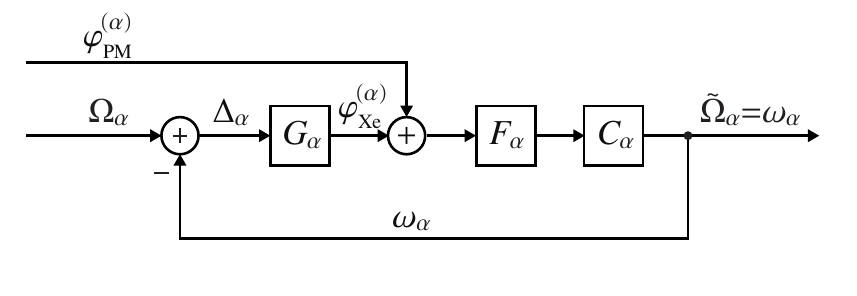}
	\caption{The block diagram of the close-loop NMR system. The input signal of the NMR system $\Delta_{\alpha}$ is the difference between the Larmor frequency $\Omega_{\alpha}$ of the nuclear spins 
	and the driving field frequency $\omega_{\alpha}$. 
	The transfer function of the NMR system is denoted by $G_{\alpha}$, which relates the output phase $\varphi_{\rm Xe}^{(\alpha)}$ to the frequency detuning $\Delta_{\alpha}$.
	The output phase $\varphi_{\rm Xe}^{(\alpha)}$, together with the Rb-PM phase noise $\varphi_{\rm PM}^{(\alpha)}$, is fed into the low-pass filter $F_{\alpha}$ to suppress high-frequency noise. 
	The PI controller $C_{\alpha}$ continuously adjusts the driving field frequency to maintain resonance of the NMR system.
	The driving frequency $\omega_{\alpha}$ is regarded as the output frequency $\tilde{\Omega}_{\alpha}$ of the PLL.}
	\label{fig:TF_digram}
\end{figure}

The spin precession frequency is obtained from the measured NMR phase $\varphi_{\rm Xe}^{(\alpha)}$ by a phase-lock loop (PLL).
As shown in Fig.~\ref{fig:TF_digram}, the measured phase $\varphi^{(\alpha)}_{\rm Xe}$, followed by a low-pass filter, is converted to a frequency shift signal by a PI controller.
The shifted frequency signal is then fed back to the driving field, so that the phase $\varphi^{(\alpha)}_{\rm Xe}$ is locked at a preset value.
Meanwhile, the shifted driving frequency $\omega_{\alpha}$ is regarded as the output frequency $\tilde{\Omega}_{\alpha}$ of the PLL, i.e., $\tilde{\Omega}_{\alpha} = \omega_{\alpha}$.
The closed-loop transfer function from the input Larmor frequency $\Omega_{\alpha}$ to the output frequency $\tilde{\Omega}_{\alpha}$ is 
\begin{equation}
	H_{\alpha}(s) = \frac{G_{\alpha}(s)C_{\alpha}(s)F(s)}{1 +G_{\alpha}(s)C_{\alpha}(s)F(s)},
\end{equation}
where $C_{\alpha}(s) = k^{(\alpha)}_{\rm P} + k^{(\alpha)}_{\rm I} /s$ is the transfer function of the PI controller of isotope $\alpha$ of Xe (with the proportional and integral parameters $k^{(\alpha)}_{\rm  P}$ and $k^{(\alpha)}_{\rm  I}$, respectively), 
and $F(s) = 1/(1+\tau_{\rm f}s)^{k_{\rm f}}$ is the transfer function of the $k_{\rm f}$-th order low-pass filter with time constant $\tau_{\rm f}$.

The transfer function $H_{\alpha}(s)$ describes the response of the measured spin precession frequency $\tilde{\Omega}_{\alpha}$ (the output frequency) to the Xe spin Larmor frequency $\Omega_{\alpha}$ (the input frequency).
In the following, $H_{\alpha}(s)$ is referred to as the frequency-frequency transfer function ($f$-$f$ transfer function).
To simplify the expression of $H_{\alpha}(s)$, we assume a short time constant for the filter $F(s)$ (e.g., $\tau_{\rm f}\sim 10~{\rm ms}$) such that its impact on the low frequency range (with frequency $2\pi f\ll \tau_{\rm f}^{-1}$) is negligible, i.e., $F(s)\approx 1$.
Furthermore, we choose the PI control parameters satisfy the following conditions
\begin{eqnarray}
	k^{(129)}_{\rm P} &=& k^{(131)}_{\rm P} \equiv k_{\rm P}
	\label{Eq:PICondition1}\\
	k^{(\alpha)}_{\rm I} &=& \Gamma_{\alpha}k^{(\alpha)}_{\rm P}.
	\label{Eq:PICondition2}
\end{eqnarray}
In this case, the $f$-$f$ transfer function reduces to
\begin{equation}
	H_{\alpha}(s) = \frac{k_{\rm P}}{s+k_{\rm P}} \equiv H(s).
	\label{Eq:tfH}
\end{equation}
Equation \eqref{Eq:tfH} shows that the $f$-$f$ transfer function behaves as a 1st order low-pass filter with unity-gain and bandwidth $k_{\rm P}$.
More importantly, under conditions \eqref{Eq:PICondition1} and \eqref{Eq:PICondition2}, the $f$-$f$ transfer function is isotope independent.
Two isotopes $^{129}{\rm Xe}$ and $^{131}{\rm Xe}$ have identical response to the change of the Larmor frequencies, which is essential in suppressing the magnetic field drift in the gyroscope signal. 

\subsubsection{Phase-frequency transfer function}

In additional to the input Larmor frequency $\Omega_{\alpha}$, the output NMR frequency is also affected by the phase $\varphi_{\rm{PM}}^{(\alpha)}$ introduced by the Rb-PM. 
To characterize the frequency response to the change of the measured NMR phase, we define the phase-frequency transfer function ($\varphi$-$f$ transfer function)
\begin{equation}
	P_{\alpha}(s) = \frac{C_{\alpha}(s)F(s)}{1 +G_{\alpha}(s)C_{\alpha}(s)F(s)} = \frac{k_{\rm P} \left(s+\Gamma_{\alpha}\right)}{s + k_{\rm P}},
	\label{Eq:tfP}
\end{equation}
where we have assumed $F(s) \approx 1$ and applied the conditions \eqref{Eq:PICondition1} and \eqref{Eq:PICondition2}.
In contrast to the $f$-$f$ transfer function $H(s)$, the $\varphi$-$f$ transfer function $P_{\alpha}(s)$ is, in general, isotope-dependent via the different spin relaxation rates $\Gamma_{129}$ and $\Gamma_{131}$.
\subsubsection{Gyroscope signal}
The output Xe NMR frequency $ \tilde{\Omega}_{\alpha}$ is affected by the spin Larmor frequency $\Omega_{\alpha}$ and the Rb-PM output phase $\varphi_{\rm PM}^{(\alpha)}$, as shown in Fig.~\ref{fig:TF_digram}. 
With the $f$-$f$ transfer function and the $\varphi$-$f$ transfer function, the output NMR frequency of isotope $\alpha$ is 
\begin{eqnarray}
	\tilde{\Omega}_{\alpha}(s) = H_{\alpha}(s) \Omega_{\alpha}(s) + P_{\alpha}(s) \varphi_{\mathrm{PM}}^{(\alpha)}(s).
\end{eqnarray}
The NMRG output rotation rate $\tilde{\Omega}_{\rm gyro}(s)$ is the linear combination of the output NMR frequencies
\begin{equation}
\tilde{\Omega}_{\rm gyro}(s) = \frac{\tilde{\Omega}_{129} - R  \tilde{\Omega}_{131}(s)}{1+R} \equiv \tilde{\Omega}_{\rm gyro}^{(f)}(s) + \tilde{\Omega}_{\rm gyro}^{(\varphi)}(s),
\label{Eq:GyroSig}
\end{equation}
where $R=\left\vert \gamma^{(129)}_{\rm Xe}/\gamma^{(131)}_{\rm Xe}\right\vert = 3.37340$. 
In Eq.~\eqref{Eq:GyroSig}, the NMRG output is separated into two parts, the contribution $\tilde{\Omega}_{\rm gyro}^{(f)}(s)$ of the input frequencies $\Omega_{\alpha}(s)$
\begin{equation}
\tilde{\Omega}_{\rm gyro}^{(f)}(s)  =  \frac{H_{129}(s) \Omega_{129}(s) -RH_{131}(s) \Omega_{131}(s) }{1+R}, 
\end{equation}
and the contribution $\tilde{\Omega}_{\rm gyro}^{(\varphi)}(s)$ of the Rb-PM measurement phase $\varphi_{\rm PM}^{(\alpha)}(s)$
\begin{equation}
\tilde{\Omega}_{\rm gyro}^{(\varphi)}(s)  = \frac{  P_{129}(s) \varphi_{\mathrm{PM}}^{(129)}(s) - RP_{131}(s) \varphi_{\mathrm{PM}}^{(131)}(s)}{1+R}.
\label{Eq:PhaseNoise}
\end{equation}
In the following, we will examine the noise spectrum of both frequency and phase contributions separately.
\subsection{Noise spectrum of NMRG}
\subsubsection{Frequency noise spectrum}
With the condition in Eqs.~\eqref{Eq:PICondition1} \& \eqref{Eq:PICondition2} and in the low frequency regime $2\pi f\ll k_{\rm P}$, the $f$-$f$ transfer function $H(s) \to 1$.
In this case, the frequency contribution to the NMRG output is
\begin{eqnarray}
	\tilde{\Omega}_{\rm gyro}^{( f)} &=&\frac{H (s) \left[\Omega_{129}(s) -R \Omega_{131}(s) \right]}{1+R}  \notag\\
		&=&  \Omega_{\mathrm{rot}}(s) + \bar{\gamma}_{\mathrm{Xe}} b_{A}(s), 
\end{eqnarray}
where 
\begin{equation}
	\bar{\gamma}_{\mathrm{Xe}}  = \frac{\vert \gamma_{\rm {Xe}}^{(129)}\gamma_{\rm Xe}^{(131)}\vert}{\vert \gamma_{\rm Xe}^{(129)}\vert +\vert \gamma_{\rm Xe}^{(131)}\vert} = 2.712~\mathrm{mHz/nT},
	\label{Eq:bar_gamma_xe}
\end{equation} 
and 
\begin{equation}
	b_{A}(s) = B_{A}^{(129)}(s) - B_{A}^{(131)}(s) 
\end{equation}
is the differential polarization field between the two isotopes.


The differential polarization field $b_A$ brings about an systematic error of the NMRG.
More importantly, the differential polarization field causes the drift of the NMRG output, if $b_A$ is changing with time.
In the low frequency regime, the NMRG output drift relates to the change $\delta b_A$ of the differential polarization field as
\begin{equation}
	\delta \tilde{\Omega}_{\rm gyro}^{(f)}(s) = \bar{\gamma}_{\mathrm{Xe}} \delta b_{A}(s).
	\label{Eq:Gyro_vs_bA}
\end{equation}
The power spectrum $S_{\rm gyro}^{(f)} (f)$ of the NMRG output induced by the differential polarization field $\delta b_A$ is
\begin{equation}
	S_{\rm gyro}^{(f)} (f) = \bar{\gamma}_{\mathrm{Xe}}^2 S_{b_{A}}(f),
	\label{Eq:NoiseSpectrum_bA}
\end{equation}
where $S_{b_{A}}(f)$ is the power spectrum of the differential polarization field.



\subsubsection{Phase noise spectrum}
\label{Sect:PhaseNoiseSpectrum}
Due to the inevitable disturbance, the Rb-PM output phase $\varphi_{\rm PM}^{(\alpha)}$ is fluctuating around a mean value $\bar{\varphi}_{\rm PM}^{(\alpha)}$, even if the phase of the detected signal is actually stable, i.e.,
\begin{equation}
	\varphi_{\rm PM}^{(\alpha)}(t) = \bar{\varphi}_{\rm PM}^{(\alpha)} + \delta \varphi_{\rm PM}^{(\alpha)}(t).
\end{equation}
The Rb-PM phase noise $\delta \varphi_{\rm PM}^{(\alpha)}(t)$ is further classified into two types, namely, the white noise $\delta \varphi_{\rm w}^{(\alpha)}(t)$ and the low-frequency colored noise $\delta \varphi_{\rm c}^{(\alpha)}(t)$
\begin{equation}
	\delta \varphi_{\rm PM}^{(\alpha)} (t) = \delta \varphi_{\rm w}^{(\alpha)}(t) + \delta \varphi_{\rm c}^{(\alpha)}(t).
	\label{Eq:PhaseNoiseComponents}
\end{equation}
The white noise $\delta \varphi_{\rm w}^{(\alpha)}(t)$ mainly arises from the photon shot noise of the probe beam.
The shot noise for the two isotopes is uncorrelated. 
The correlation function is
\begin{equation}
	\langle \delta \varphi_{\rm w}^{(\alpha)}(t)\delta \varphi_{\rm w}^{(\alpha')}(t')\rangle = \frac{S_{\varphi, {\rm w}}^{(\alpha)}}{2} \delta_{\alpha, \alpha'} \delta(t-t'),
	\label{Eq:WhitePhaseNoiseCorrelation}
\end{equation}
where $S_{\varphi, {\rm w}}^{(\alpha)}$ is the (one-sided) power spectrum of the white noise $\delta \varphi_{\rm w}^{(\alpha)}(t)$.

The colored phase noise $\delta \varphi_{\rm c}^{(\alpha)}(t)$ usually comes from the low-frequency drift of the system control parameters (e.g., cell temperature $T$, laser power $P$, etc.), 
which affects both Xe isotopes in the same manner.
As a result, the colored phase noise of the two isotopes are highly correlated, taking the form
\begin{equation}
	\delta \varphi_{\rm c}^{(\alpha)} (t) = \lambda_{\alpha}\delta \varphi_{\rm c}(t),
\end{equation}
with identical time-dependence $\delta \varphi_{\rm c}(t)$ but different amplitudes $\lambda_{\alpha}$.

In the frequency domain, the systematic error of the NMRG output induced by the Rb-PM phase noise is
\begin{eqnarray}
	\label{Eq:phase_noise}
	\delta \tilde{\Omega}_{\rm gyro}^{(\varphi)} (s) &=& \frac{P_{129}(s) \delta \varphi_{\mathrm{PM}}^{(129)}(s) - R P_{131}(s) \delta \varphi_{\mathrm{PM}}^{(131)}(s) }{1+R}\\
	&=&\frac{\Gamma_{129} \delta \varphi_{\mathrm{w}}^{(129)}(s) - R \Gamma_{131} \delta \varphi_{\mathrm{w}}^{(131)}(s) }{1+R}+ \xi_{\rm c} \delta \varphi_{\mathrm{c}}(s), \notag
\end{eqnarray}
where $\xi_{\rm c}$ is the suppression factor of the colored noise
\begin{equation}
	\xi_{\mathrm{c}} \equiv \frac{\lambda_{129}\Gamma_{129} - R\lambda_{131}\Gamma_{131}}{1+R}.
	\label{Eq:suppresionFactor}
\end{equation}
In the second line of Eq.~\eqref{Eq:phase_noise}, we have applied the fact that the $\varphi$-$f$ transfer function approaches a constant $P_{\alpha}(s) \to \Gamma_{\alpha}$ in the low frequency regime $\vert s\vert \ll \Gamma_{\alpha}$.

Furthermore, it is reasonable to assume the white noise and the colored noise are uncorrelated, i.e., 
\begin{equation}
	\langle \delta\varphi_{ {\rm w}}^{(\alpha)}(t)  \delta\varphi_{\rm c}(t')\rangle = 0.
	\label{Eq:PhaseNoiseCrossCorrelation}
\end{equation}
With the power spectrum of the colored noise $\delta \varphi_{\rm c}(t)$ denoted by $S_{\varphi, {\rm c}}(f)$, the power spectrum induced by the phase noise is
\begin{equation}
	S_{\rm gyro}^{(\varphi)}(f) = \left(\frac{\Gamma_{129}}{1+R}\right)^2 S_{\varphi,\mathrm{w}}^{(129)} +  \left(\frac{R\Gamma_{131}}{1+R}\right)^2 S_{\varphi,\mathrm{w}}^{(131)} + \xi_{\mathrm{c}}^2 S_{\varphi,\mathrm{c}}(f).
\label{Eq:phase_spectrum}
\end{equation}

The first two terms of Eq.~\eqref{Eq:phase_spectrum} are the white noise mainly induced by the shot-noise of the probe beam, which determine the sensitivity of the NMRG.
The phase noise spectrum $S_{\varphi, {\rm c}}(f)$ [the 3rd term of Eq.~\eqref{Eq:phase_spectrum}] 
and the noise spectrum $S_{b_A}(f)$ in Eq.~\eqref{Eq:NoiseSpectrum_bA} both contain the $1/f$ component, which corresponds to the bias instability of the NMRG. 
In the following, we will study the physical origin of the colored phase noise $\delta \varphi_{\rm c}^{(\alpha)}(t)$, and show that the contribution of phase noise can be significantly suppressed by the self-calibrating method.
\section{NMRG Noise: Parametric Magnetometer and Phase Measurement Noise}
\label{Sect:PM}
In this section, we start from the equation of motion of Rb atomic spin polarization $\langle \mathbf{S}\rangle$ under a parametric modulation of the Larmor frequency,
and study the physical origin of the colored phase noise $\delta \varphi_{\rm c}^{(\alpha)}(t)$.
The dynamics of the transverse component $\langle S_+\rangle=\langle S_x\rangle + i\langle S_y\rangle$ is governed by 
\begin{eqnarray}
	\frac{d \langle S_+\rangle }{dt} &=&\left[-i\Omega_0-i\gamma_{\rm Rb}B_{\rm ac}\cos( \omega_0 t+\theta_{\rm ac})  -\Gamma_{\rm Rb}\right]\langle S_+\rangle\notag\\
	&&+i\gamma_{\rm Rb}b_{\rm c}(t)\langle S_z\rangle,
	\label{Eq:BlochEq}
\end{eqnarray}
where $\gamma_{\rm Rb}$ is the gyromagnetic ratio of Rb atom, $\Omega_0 = \gamma_{\rm Rb}B_0$ is the Larmor frequency of Rb atomic spins in a static magnetic field $B_0$,
$B_{\rm ac}$, $\omega_0$ and $\theta_{\rm ac}$ are the amplitude, frequency and phase of the modulation magnetic field along the $z$ direction, 
$\Gamma_{\rm Rb}$ is the transverse spin relaxation rate of Rb,  and $b_{\rm c} (t) = b_x(t) + i b_y(t)$ is the complex magnetic signal to be measured.
For the harmonic oscillating fields in the $x$ and $y$ directions with a given frequency $\omega$, the complex signal $b_{\rm c}(t)$ is 
\begin{eqnarray}
	b_{\rm c}(t) &=& b_{x0} \cos(\omega t+  \phi_x) + i b_{y0} \cos(\omega t +\phi_y)\notag\\
	 & \equiv & b^{+} e^{i\omega t} + b^{-} e^{-i\omega t},
	 \label{Eq:ComplexSignal}
\end{eqnarray}
where $b_{x0/y0}$ and $\phi_{x/y}$ are the amplitude and phase of the oscillation in the $x/y$ direction,
and $b^{\pm} = \left[b_{x0}\exp(\pm i\phi_x) +i b_{y0} \exp(\pm i\phi_y)\right]/2$ are the complex amplitudes of the positive/negative frequency components.
Particularly, with equal amplitudes $b_{x0}=b_{y0} = b_0$, the positive frequency component $b^{+}\exp(i\omega t)$ corresponds to the left-hand circularly polarized (LCP) field with $\phi_{x}-\phi_{y}=\pi/2$,
while the negative frequency component $b^{-}\exp(-i\omega t)$ corresponds to the right-hand circularly polarized (RCP) field with $\phi_{x}-\phi_{y}=-\pi/2$.

\subsection{Solution of Rb-PM equation of motion}
\subsubsection{Adiabatic and narrow-linewidth approximation}
\label{Sect:AdiabaticNarrowLinewidthApproximation}
In the NMRG system, the oscillation frequency $\omega$ of the signal $b_{\rm c}(t)$ is usually much smaller than spin relaxation rate of Rb atomic spins, i.e., $\omega \ll \Gamma_{\rm Rb}$.
To the lowest order approximation (the adiabatic approximation), the time-dependence of the signal $b_{\rm c}(t)$ is ignored when solving Eq.~\eqref{Eq:BlochEq} within the Rb atomic spin relaxation time scale $\sim \Gamma_{\rm Rb}^{-1}$. 
The solution in this adiabatic limit has been extensively studied \cite{Walker2016,eklundMicrogyroscopeBasedSpinPolarized2008, tang2019optimizations}.
For example, with the narrow linewidth condition $\Gamma_{\rm Rb} \ll \gamma_{\rm Rb} B_0$, the solution of the $x$ component reads
\begin{eqnarray}
	\langle S_x(t)\rangle &=& \frac{\gamma_{\rm Rb}\langle S_z\rangle J_{1}}{\Gamma_{\rm Rb}}\left[ -(J_0 - J_2)b_{x}(t)\sin(\omega_0 t+\theta_{\rm ac}) \right.\notag\\
	&&\left.  + (J_0 + J_2)b_{y}(t)\cos(\omega_0 t+\theta_{\rm ac})\right],
	\label{Eq:StaticSolution}
\end{eqnarray}
where $J_{n} \equiv J_{n}(\eta)$ is the $n$th order Bessel function evaluated at the modulation strength parameter $\eta = \gamma_{\rm Rb} B_{\rm ac}/\omega_0$.
The transverse fields $b_x(t)$ and $b_y(t)$ are extracted by demodulating $\langle S_x(t)\rangle$
\begin{align}
	&C^{(1)}(t) = \overline{ \langle S_x(t)\rangle \cdot V_{\rm ref}^{(1)}(t) } \notag\\
		&= \frac{\gamma_{\rm Rb}\langle S_z\rangle J_{1}}{\sqrt{2}\Gamma_{\rm Rb}}\left[ i(J_0 - J_2) b_{x}(t) + (J_0 + J_2)b_{y}(t)\right]e^{i(\theta_{\rm ac}-\theta_1)},
	\label{Eq:StaticSolution_1stDemod}
\end{align}
where $V_{\rm ref}^{(1)} \equiv \sqrt{2}\exp\left[-i( \omega_0 t + \theta_1)\right]$ is the reference signal with frequency $\omega_0$ and demodulation phase $\theta_1$, 
and $\overline{x(t)}$ stands for the low-pass filter operation on a signal $x(t)$, which keeps the near-DC component with frequency $f\ll \omega_0/2\pi$.
The demodulation with reference $V_{\rm ref}^{(1)}$ in Eq.~\eqref{Eq:StaticSolution_1stDemod} is referred to as the {\it first-demodulation} hereafter.
The quadratures of the first-demodulation are oscillating signals with the same frequency $\omega$ as the fields $b_x(t)$ and $b_y(t)$. 
For example, for a given demodulation phase $\theta_1$, the quadrature $X^{(1)}(t; \theta_1) = \Re [C^{(1)}(t; \theta_1)]$ is
\begin{eqnarray}
	X^{(1)}(t) &=& \frac{\gamma_{\rm Rb}\langle S_z\rangle J_{1}}{\sqrt{2}\Gamma_{\rm Rb}}\left[ (J_2 - J_0)\sin(\theta_{\rm ac}-\theta_1) b_{x}(t) \right.\notag\\
	&& \left. + (J_0 + J_2)\cos(\theta_{\rm ac}-\theta_1) b_{y}(t)\right].
	\label{Eq:StaticQuadratures1}
\end{eqnarray}

A {\it second-demodulation} is performed by demodulating the quadrature $X^{(1)}(t; \theta_1)$ with the reference signal $V_{\rm ref}^{(2)}(t)=\sqrt{2}\exp\left[-i (\omega t + \theta_2)\right]$.
The complex demodulation output is
\begin{eqnarray}
	C^{(2)} &=& \overline{X^{(1)}(t) \cdot V_{\rm ref}^{(2)}(t)}\notag \\
	&=& \frac{\gamma_{\rm Rb}\langle S_z\rangle J_1}{2\Gamma_{\rm Rb}}\left[ (J_2-J_0)\sin(\theta_{\rm ac} -\theta_1) b_{x0}e^{i(\phi_x - \theta_2)}\right.\notag\\
	&& + \left. (J_0+J_2)\cos(\theta_{\rm ac} -\theta_1) b_{y0}e^{i(\phi_y - \theta_2)}\right].
	\label{Eq:StaticSolution_2ndDemod}
\end{eqnarray}
The solution \eqref{Eq:StaticSolution} and the two-stage demodulation in Eqs.~\eqref{Eq:StaticSolution_1stDemod}-\eqref{Eq:StaticSolution_2ndDemod} provide a good description of measurement principle of the fields $b_x(t)$ and $b_y(t)$,
if the accuracy requirement is not too high. 

Unfortunately, the results above are inadequate in analysing the NMRG stability.
At least two important factors must be considered to establish a quantitatively accurate theoretic model.
Firstly, the narrow linewidth condition $\Gamma_{\rm Rb}\ll \gamma_{\rm Rb}B_0$ is not always satisfied.
The finite linewidth correction was discussed previously\cite{tang2019optimizations}.
Secondly, the adiabatic approximation is not precise enough to explain the observed phase of $b_x(t)$ and $b_y(t)$. 
The NMRG application requires the phase measurement with a high accuracy in the order of $\sim 10^{-3}~{\rm deg}$. 
An exact solution to Eq.~\eqref{Eq:ComplexSignal} is necessary in analysing the phase measurement stability.

\subsubsection{Exact solution}
The exact solution to Eq.~\eqref{Eq:BlochEq} is presented in Appendix~\ref{Appendix:Rb-PM Response}. 
The spin components $\langle S_x(t)\rangle$ is
\begin{eqnarray}
	&&\langle S_x(t)\rangle
	=\frac{\gamma_{\rm Rb} \langle S_z\rangle}{2\Gamma_{\rm Rb}} \notag\\
	&\times &\sum_{p=-\infty}^{\infty}
	\left(\mathcal{A}_p^{+}b^{+}e^{i \omega t}+\mathcal{A}_p^{-}b^{-}e^{ -i\omega t}\right)e^{ip(\omega_0 t+\theta_{\rm ac})}+ {\rm c.c.},
	\label{Eq:ExactSolution}
\end{eqnarray}
where the complex amplitudes $\mathcal{A}_p^{\pm}$ are
\begin{equation}
	\mathcal{A}_p^{\pm}(\Omega_0, \eta, \omega_0, \Gamma_{\rm Rb}, \omega) = \sum_{n=-\infty}^{\infty} \frac{ \Gamma_{\rm Rb} J_{n-p}(\eta) J_n(\eta) }{  \Omega_0+n\omega_0  \pm \omega -i\Gamma_{\rm Rb}}.
	\label{Eq:ComplexAmplitude_A_p}
\end{equation}
Notice that the complex amplitudes $\mathcal{A}_p^{\pm}$ depend on the frequency $\omega$ of the signal to be measured, which is the consequence of going beyond the adiabatic approximation.

Similar to Eqs.~\eqref{Eq:StaticSolution_1stDemod}, the $\langle S_x(t)\rangle$ is demodulated with the $p$th order reference signal $V_{{\rm ref}, p}^{(1)} = \sqrt{2}\exp[-i (p \omega_0 t + \theta_1)]$,
i.e., $C^{(1)}_p(t) = \overline{ \langle S_x(t)\rangle \cdot V_{{\rm ref}, p}^{(1)}(t) }$, 
and the quadrature $X_p^{(1)}(t) = \Re [C_p^{(1)}(t)]$ of the first-demodulation is a harmonic oscillating signal of frequency $\omega$
\begin{eqnarray}
	X_p^{(1)}(t) &=& \frac{\sqrt{2}\gamma_{\rm Rb} \langle S_z\rangle }{4\Gamma_{\rm Rb}}\left[\left(b^{+}\mathcal{A}_{p}^{+}+b^{-*}\mathcal{A}_{-p}^{-*}\right)e^{i(p\theta_{\rm ac}-\theta_1) }\right.\notag\\
	&&\left.+\left(b^{+}\mathcal{A}_{-p}^{+}+ b^{-*}\mathcal{A}_{p}^{-*} \right)e^{-i(p\theta_{\rm ac}-\theta_1) } \right]e^{ i\omega t}+ {\rm c.c.}.
\end{eqnarray}
The second-demodulation of the signal $X_p^{(1)}(t)$ results in the complex amplitude
$C_p^{(2)} =\overline{X_p^{(1)}(t) \cdot V_{\rm ref}^{(2)}(t)}$ as 
\begin{equation}
	C_p^{(2)} =\frac{\gamma_{\rm Rb} \langle S_z\rangle }{4\Gamma_{\rm Rb}}\left[G_{p}^{(x)}(\theta_1)b_{x0}e^{i(\phi_x- \theta_2)} - G_{p}^{(y)}(\theta_1)b_{y0}e^{i( \phi_y-\theta_2)}\right],
	\label{Eq:SecondDemodulationExaxt}
\end{equation}
where $G_p^{(x)}(\theta_1)$ and $G_p^{(y)}(\theta_1)$ are the dimensionless gain functions of the Rb-PM, and they relate to the complex amplitude $\mathcal{A}_p^{\pm}$ as
\begin{eqnarray}
	G^{(x)}_{p}(\theta_1) &=&(\mathcal{A}^{+}_{p}+\mathcal{A}^{-*}_{-p}) e^{i(p\theta_{\rm ac} -\theta_1)} \notag\\
	       		&&+ (\mathcal{A}^{+}_{-p}+\mathcal{A}^{-*}_{p}) e^{-i(p\theta_{\rm ac} -\theta_1)},\label{Eq:Gx}\\
	G^{(y)}_{p}(\theta_1) &=& -i(\mathcal{A}^{+}_{p}-\mathcal{A}^{-*}_{-p}) e^{i(p\theta_{\rm ac} -\theta_1)} \notag\\
	  		&&- i (\mathcal{A}^{+}_{-p}-\mathcal{A}^{-*}_{p}) e^{-i(p\theta_{\rm ac} -\theta_1)}.\label{Eq:Gy}
\end{eqnarray}
Furthermore, with the amplitude ratio $\chi = b_{y0}/b_{x0}$ and the relative phase $\phi = \phi_x - \phi_y$ of the field $b_{\rm c}(t)$ to be measured,
Eq.~\eqref{Eq:SecondDemodulationExaxt} is simplified as
\begin{equation}
	C_{p}^{(2)} = \frac{\gamma_{\rm Rb} \langle S_z\rangle }{4\Gamma_{\rm Rb}} G_p(\theta_1; \chi, \phi) b_{x0} e^{i(\phi_x-\theta_2)},
\end{equation}
where
\begin{equation}
	G_p(\theta_1;\chi, \phi) =  G_p^{(x)}(\theta_1) -\chi e^{-i\phi } G_p^{(y)}(\theta_1).\label{Eq:Gpm}
\end{equation}
Particularly, with $\chi=1$ and $\phi = \pm \pi/2$, the gain functions for the LCP ($+$) and RCP ($-$) fields  become
\begin{equation}
	G_p^{(\pm)}(\theta_1) =  G_p^{(x)}(\theta_1)  \pm i G_p^{(y)}(\theta_1).\label{Eq:Gpm1}
\end{equation}

In general, the function $G_p(\theta_1; \chi, \phi)$ is complex-valued, whose magnitude $\vert G_p(\theta_1; \chi, \phi)\vert$ is the gain coefficient of the Rb-PM output for a given input transverse field.
The phase angle 
\begin{equation}
	\varphi_{\rm PM} = \arg[G_p(\theta_1; \chi, \phi)]
\end{equation}
describes the additional phase introduced by the Rb-PM.
A constant phase, in principle, can be ignored in the NMRG application.
However, as shown in the following sections, the Rb-PM phase $\varphi_{\rm PM}$ depends on several control parameters of the NMRG system, which drift inevitably.
The phase $\varphi_{\rm PM}$ of the gain function $G_p(\theta_1; \chi, \phi)$ is essential in the analysis of the NMRG stability.

\subsection{Gain functions and working point}
Before discussing the properties of the gain functions, we present the key parameters in a typical NMRG system. 
The Larmor frequency of Rb atomic spins in the static field $B_0$ is $\Omega_0\sim 2\pi \times 100~{\rm kHz}$, the Rb spin relaxation rate is $\Gamma_{\rm Rb} \sim 2\pi \times 7.5~{\rm kHz}$, and the NMR signal frequency to be measured is $\omega \sim 2\pi \times 100~{\rm Hz}$.
The frequency $\omega_0$ of the parametric driving field differs to the Rb Larmor frequency by a detuning $\Delta_{\rm Rb} \equiv \Omega_0 - \omega_0$, which is usually small compared to the Rb spin relaxation rate, i.e., $\vert \Delta_{\rm Rb}\vert \ll \Gamma_{\rm Rb}$.

Normalizing the frequencies $\Omega_0$, $\omega$ and $\Delta_{\rm Rb}$ by the Rb spin relaxation rate $\Gamma_{\rm Rb}$, we can define the following dimensionless quantities: $\zeta = \omega_0/\Gamma_{\rm Rb}\gg 1$, $x=\omega/\Gamma_{\rm Rb}\ll 1$, and $\delta = \Delta_{\rm Rb} /\Gamma_{\rm Rb}\ll 1$.
With these parameters, the properties of the gain functions are analysed by the expanding $G_p^{(x/y)}(\theta_1)$ to the leading order of the small quantities $\zeta^{-1}$, $x$ and $\delta$.
Up to the linear order, the magnitudes of the gain functions are 
\begin{eqnarray}
	\vert G^{(x)}_p (\theta_1) \vert &=& \left \vert d_p(\eta)  \sin(\Theta_p^{(x)} - \theta_1 )\right \vert + \mathcal{O}_2,
	\label{Eq:GainX}\\
	\vert G^{(y)}_p (\theta_1) \vert &=& \left \vert s_p(\eta)  \sin(\Theta_p^{(y)} - \theta_1 )\right \vert + \mathcal{O}_2,
	\label{Eq:GainY}
\end{eqnarray}
where $\mathcal{O}_2$ stands for second-order correction terms, 
and the phase shift $\Theta_p^{(x)}$ and $\Theta_p^{(y)}$ are 
\begin{eqnarray}
	\Theta_p^{(x)} &=& p\theta_{\rm ac} + k_p(\eta)\cdot\delta+k_p'(\eta)\cdot\zeta^{-1},
	\label{Eq:WP_PhaseX}\\
	\Theta_p^{(y)} &=& p\theta_{\rm ac} + q_p(\eta)\cdot\delta+q_p'(\eta)\cdot\zeta^{-1} + \frac{\pi}{2}.
	\label{Eq:WP_PhaseY}
\end{eqnarray}
In Eqs.~\eqref{Eq:GainX}-\eqref{Eq:WP_PhaseY}, the functions of the modulation strength $\eta$ are defined as
\begin{eqnarray}
	s_p(\eta) &\equiv& 2J_{-1}(\eta)\left[J_{p-1}(\eta) + J_{-p-1}(\eta)\right],
	\label{Eq:s_p}\\
	d_p(\eta) &\equiv& 2J_{-1}(\eta)\left[J_{p-1}(\eta) - J_{-p-1}(\eta)\right],
	\label{Eq:d_p}\\
	s_p'(\eta) &\equiv& 2\sum_{n\neq 0} n^{-1}J_{n-1}(\eta)\left[J_{n+p-1}(\eta)+J_{n-p-1}(\eta)\right],\\
	d_p'(\eta) &\equiv& 2\sum_{n\neq 0} n^{-1}J_{n-1}(\eta)\left[J_{n+p-1}(\eta)-J_{n-p-1}(\eta)\right],\\
	k_p(\eta) &\equiv& \frac{s_p(\eta)}{d_p(\eta)}, \quad k_p'(\eta) = \frac{s_p'(\eta)}{d_p(\eta)},
	\label{Eq:k_p}\\
	q_p(\eta) &\equiv& \frac{d_p(\eta)}{s_p(\eta)}, \quad q_p'(\eta) = \frac{d_p'(\eta)}{s_p(\eta)}.
	\label{Eq:q_p}
\end{eqnarray}
\begin{figure}
	\centering
	\includegraphics[scale=1]{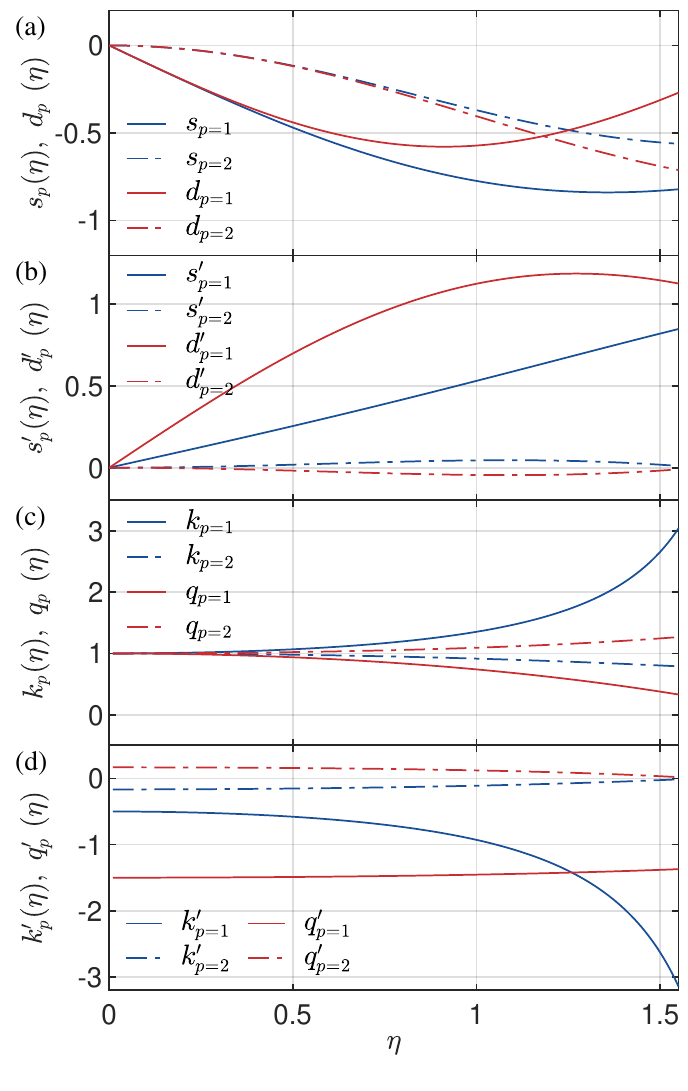}
	\caption{Numerical values of $s_p(\eta)$, $d_p(\eta)$, $s'_p(\eta)$, $d'_p(\eta)$, $k_p(\eta)$, $k'_p(\eta)$, $q_p(\eta)$, and $q'_p(\eta)$, which are defined in Eqs.~\eqref{Eq:s_p} - \eqref{Eq:q_p}.}  
	\label{fig:CalPhase}
\end{figure}

The magnitude $\vert G_p^{(x/y)}(\theta_1)\vert$ is minimized when $\theta_1 = \Theta_p^{(x/y)}$.
Up to the second order correction, the $b_x(t)$ and $b_y(t)$ fields can be extracted independently by choosing proper first-demodulation phases $\theta_1 = \Theta_p^{(x)}$ or $\Theta_p^{(y)}$.
In the limit $\delta \to 0$ and $\zeta \to \infty$,  $\Theta_p^{(y)}=\Theta_p^{(x)}+\pi/2$. 
In this case, Eq.~\eqref{Eq:GainX} and \eqref{Eq:GainY} are reduced to the simple case in Eq.~\eqref{Eq:StaticSolution_2ndDemod}.

Distinguishing $b_x(t)$ and $b_y(t)$ signals with proper demodulation phase $\theta_1$ is useful in the NMRG application.
As discussed in Sect.~\ref{Sect:TFNMRG}, the nuclear spins in NMRG are driven by transverse AC fields. 
In general, the signal detected by the Rb-PM is a combination of the driving field and the true NMR signal from the nuclear spins.
To remove the contribution of the AC driving field applied along the $y$ direction, we choose the demodulation phase $\theta_1=\Theta_p^{(y)}$.
In the following, unless stated otherwise, our Rb-PM is always working with the demodulation phase $\theta_1 = \Theta_p^{(y)}$, which is referred to as the {\it working point} (WP) of the Rb-PM.

A detailed calculation of gain functions $G_p(\theta_1; \chi, \phi)$ near the WP $\theta_1 \approx \Theta_p^{(y)}$ are presented in the Appendix~\ref{Appendix:Gainfunctions}.
Since the fields created by the precessing nuclear spins are circularly polarized, the behaviour of the LCP and RCP gain functions $G_p^{(\pm)}(\theta_1)$ is crucial in analysing the NMRG stability. 
According to Eqs.~\eqref{Eq:GainFunctionLCPPhase} - \eqref{Eq:GainFunctionRCPPhaseWP} and the discussions in Appendix~\ref{Appendix:Gainfunctions}, the phase angle of $G_p^{(\pm)}(\theta_1)$ near the WP is
\begin{equation}
	\varphi_{\rm PM}(\theta_1) = \arg[G_p^{(\pm)}(\theta_1)] = \pm k_p(\eta) \left(\theta_1 -\Theta_p^{(y)}\right) - \frac{\omega}{\Gamma_{\rm Rb}}.
	\label{Eq:PhaseAngle_Gpm}
\end{equation}
The Rb-PM phase $\varphi_{\rm PM}$ depends on various control parameters via the WP phase $\Theta_p^{(y)}$ in Eq.~\eqref{Eq:WP_PhaseY}, 
including the DC magnetic field $B_0$, the amplitude and phase of the parametric modulation field $\eta$ and $\theta_{\rm ac}$, 
the Rb spins relaxation rate $\Gamma_{\rm Rb}$, and the frequency $\omega$ of the field to be detected.

The drift of these control parameters causes the Rb-PM phase noise.
Assume that, at an initial time $t_0$, the phase angle is $\varphi_{\rm PM}(\theta_1; t_0) = \arg[G_p^{(\pm)}(\theta_1; t_0)]$. 
According to Eq.~\eqref{Eq:PhaseAngle_Gpm}, the change of the phase of the parametric driving field $\theta_{\rm ac}(t) = \theta_{\rm ac}(t_0) + \delta \theta_{\rm ac}$, 
the DC magnetic field $B_0(t) = B_0(t_0) + \delta B_0$ and the Rb spin relaxation $\Gamma_{\rm Rb}(t) = \Gamma_{\rm Rb}(t_0) + \delta \Gamma_{\rm Rb}(t)$
brings the Rb-PM phase to a new value $\varphi_{\rm PM}(\theta_1; t) = \arg[G_p^{(\pm)}(\theta_1; t)]$ at a later time $t$. 
Then, the phase noise is $\delta \varphi_{\rm PM}(t) = \varphi_{\rm PM}(t) - \varphi_{\rm PM}(t_0)$, 
and attributed to the small variations $\delta \theta_{\rm ac}$, $\delta B_0$ and $\delta \Gamma_{\rm Rb}$ as
\begin{eqnarray}
	\delta \varphi_{\rm PM}(t)  &=& \mp pk_p(\eta) \delta \theta_{\rm ac} \mp \frac{\gamma_{\rm Rb}}{\Gamma_{\rm Rb}} \delta B_0 + \chi_p^{(\pm)} \frac{\delta \Gamma_{\rm Rb}}{\Gamma_{\rm Rb}},\label{Eq:PhaseVariance_pm}
\end{eqnarray}
where the coefficient $\chi_p^{(\pm)}$ is 
\begin{eqnarray}
	\chi_p^{(\pm)} &=& \frac{\pm \Delta_{\rm Rb} + \omega}{\Gamma_{\rm Rb}}  \mp \frac{d'_p(\eta)}{d_p(\eta)}\frac{\Gamma_{\rm Rb}}{\omega_0}.  \label{Eq:Chi_pm}
\end{eqnarray}
Equations~\eqref{Eq:PhaseVariance_pm}-\eqref{Eq:Chi_pm} demonstrate the quantitative relation between the phase noise and the control parameters.
Notice that the phase noise induced by the change $\delta \eta(t)$ of the modulation amplitude is a small quantity of second order, if the Rb-PM is operating near the WP, i.e., $\theta_1 \approx \Theta_p^{(y)}$.

\begin{figure}[t]
	\centering
	\includegraphics[scale=1]{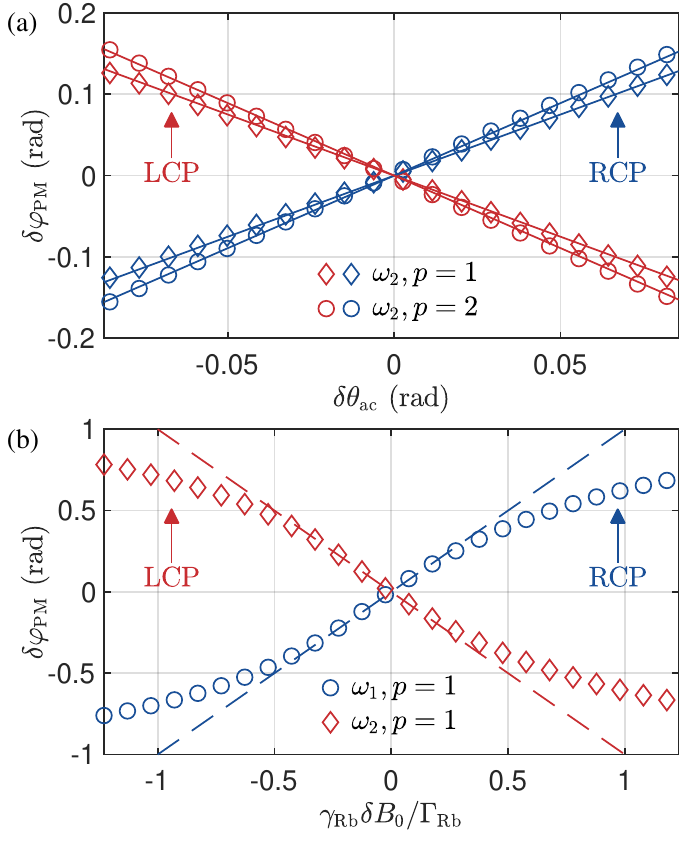}
	\caption{The variation of the Rb-PM phase $\delta \varphi_{\rm PM}$ induced by the change of (a) AC modulation phase $\delta \theta_{\rm ac}$ and (b) DC magnetic field $\delta B_0$.
	Symbols are measured data, and straight lines are theoretical results according to Eq.~\eqref{Eq:PhaseVariance_pm}.}
	\label{fig:OLCalPhase}
\end{figure}

To verify the dependence in Eq.~\eqref{Eq:PhaseVariance_pm}, 
we generate an LCP or RCP {\it calibration signal} with a stable frequency $\omega_{\rm cal}$ far from the Xe resonance frequencies, 
and measure the Rb-PM output phase $\varphi_{\rm PM}^{(\rm cal)}$.
Figure~\ref{fig:OLCalPhase} presents the Rb-PM phase change induced by sweeping the AC modulation phase $\delta \theta_{\rm ac}$ and the DC magnetic field $\delta B_0$ around the WP.
The measured data is in good agreement with the first two terms of Eq.~\eqref{Eq:PhaseVariance_pm}.

\begin{figure}[tbh]
	\centering
	\includegraphics[scale=1]{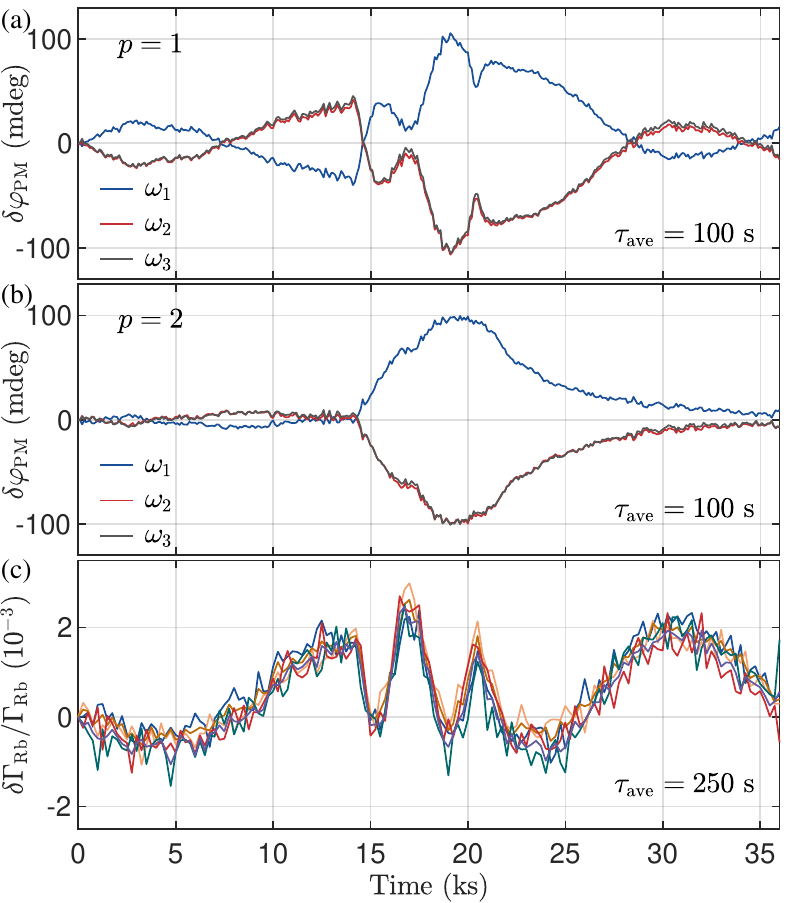}
	\caption{
	(a) Rb-PM phase variation $\delta \varphi_{\rm PM}(t)$ of order $p=1$ for three calibration signals with frequencies $\omega_1 = 2\pi\times45~{\rm Hz}$ (RCP), $\omega_2 = 2\pi\times140~{\rm Hz}$ (LCP) and $\omega_3 = 2\pi\times330~{\rm Hz}$ (LCP).
	Each data point represents a $\tau_{\rm ave} = 100~{\rm s}$ moving average of the raw data.
	(b) The same as (a) but for order $p=2$.
	(c) The normalized phase difference/sum of the phases shown in (a) and (b). All the six curves (for different frequencies and $p=1$ and $2$) coincide showing that the relative change of the Rb spin relaxation rate $\delta \Gamma_{\rm Rb}/\Gamma_{\rm Rb}$ agrees well with Eqs.~\eqref{Eq:ComparePhase1} \& \eqref{Eq:ComparePhase2}.
	Each data point represents a $\tau_{\rm ave} = 250~{\rm s}$ moving average of the raw data.
	}
	\label{fig:PMOpenLoop}
\end{figure}

The verification of the contribution of $\delta \Gamma_{\rm Rb}$ [the third term of Eq.~\eqref{Eq:PhaseVariance_pm}] to $\delta \varphi_{\rm PM}$ is not straightforward.
Although the Rb spin relaxation $\Gamma_{\rm Rb}$ can be changed by, e.g., varying cell temperature or the optical pumping rate, these operations will also affect the overall state of the Rb-PM.
Solely changing $\Gamma_{\rm Rb}$ while leaving all the other parameters unchanged is difficult.
To solve this problem, we generate several calibration signals with different frequencies $\omega_{k}$, 
set the demodulation phase $\theta_1$ at the WP at time $t_0$ [i.e., $\theta_1=\Theta_p^{(y)}(t_0)$], 
and measure the change of the Rb-PM output phases $\left .\varphi^{\rm (cal)}_{\rm PM}(t)\right \vert_{\omega_{k}}$ as functions of time $t$.
According to Eq.~\eqref{Eq:PhaseVariance_pm}, for two calibration signal with the same polarity (both LCP or RCP), 
the Rb-PM output phase difference is
\begin{equation}
	\left .\varphi_{\rm PM}^{(\rm cal)}\right\vert_{\omega_1} - \left . \varphi_{\rm PM}^{(\rm cal)}\right\vert_{\omega_2} = \frac{\omega_1-\omega_2}{\Gamma_{\rm Rb}} \frac{\delta \Gamma_{\rm Rb}}{\Gamma_{\rm Rb}},
	\label{Eq:ComparePhase1}
\end{equation}
and the sum of the Rb-PM output phases with opposite polarity (an LCP and an RCP) is
\begin{equation}
	\left .\varphi_{\rm PM}^{(\rm cal)}\right\vert_{\omega_1} + \left . \varphi_{\rm PM}^{(\rm cal)}\right\vert_{\omega_2} = \frac{\omega_1+\omega_2}{\Gamma_{\rm Rb}} \frac{\delta \Gamma_{\rm Rb}}{\Gamma_{\rm Rb}}.
	\label{Eq:ComparePhase2}
\end{equation}
In either case, the relative change of the Rb spin relaxation rate can be obtained by normalizing the phase difference or sum with $(\omega_1 - \omega_2)/\Gamma_{\rm Rb}$ or $(\omega_1 + \omega_2)/\Gamma_{\rm Rb}$.
Figure~\ref{fig:PMOpenLoop} shows the measured Rb-PM phase and the normalized phase difference/sum of three different calibration frequencies.
With the normalization, all the resultant phase difference/sum coincide, showing that the relative change of the Rb spin relaxation rate $\delta \Gamma_{\rm Rb} / \Gamma_{\rm Rb} \sim 10^{-3}$ in our NMRG system.

According to Eqs.~\eqref{Eq:PhaseVariance_pm} \& \eqref{Eq:Chi_pm}, with $\chi_{p=1}^{(\pm)}\sim 0.1$, 
the relative change $\delta \Gamma_{\rm Rb} / \Gamma_{\rm Rb}$ contributes $\sim 10~{\rm mdeg}$ to the total Rb-PM phase change.
As the measured Rb-PM phase change is in the order of $100~{\rm mdeg}$, the change of the AC modulation phase $\delta \theta_{\rm ac}$ 
and the change of the DC magnetic field $\delta B_0$ [the first two terms of Eq.~\eqref{Eq:PhaseVariance_pm}] are the main sources of Rb-PM phase instability.
To eliminate the systematic error induced by $\delta \theta_{\rm ac}$ and $\delta B_0$, we develop the self-calibrating method, as discussed in the following.

\subsection{Self-calibrating Rb-PM}
\begin{figure}[tbh]
	\centering
	\includegraphics[scale=1]{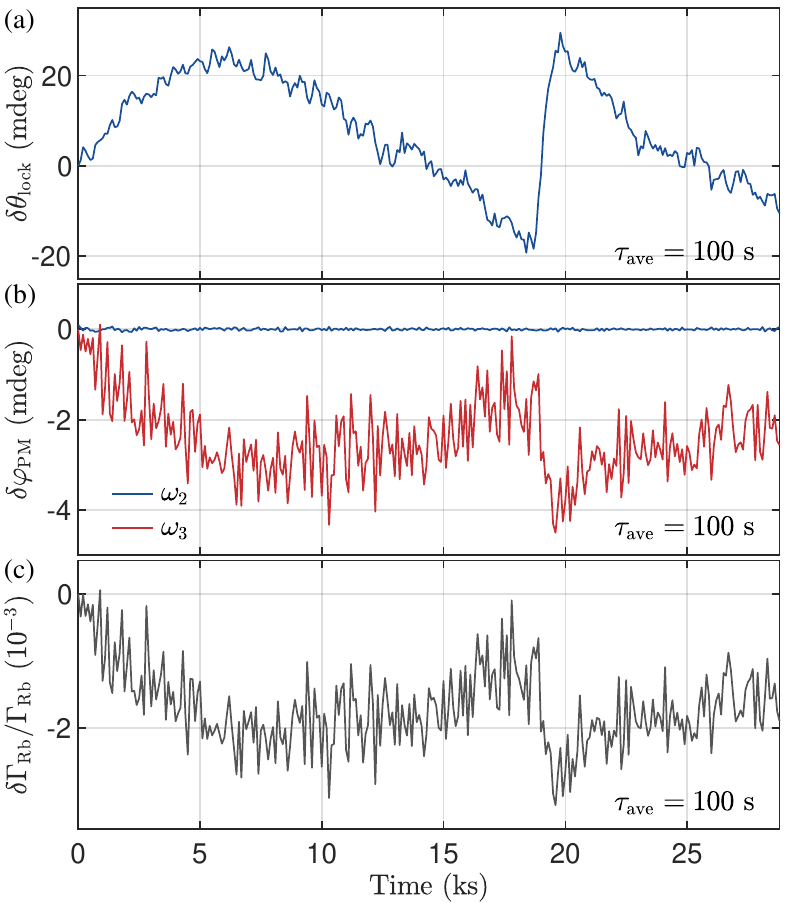}
	\caption{ Implementation of the self-calibrating Rb-PM.
	(a) The feed-back quantity $\delta\theta_{\rm lock}(t)$ of the demodulation phase, i.e. $\delta\theta_{\rm lock}(t) = \theta_{\rm lock}(t)-\theta_{\rm lock}(t_0)$ with $\theta_{\rm lock}(t_0)=\Theta_p^{(y)}(t_0)$, for a LCP calibration signal with frequency $\omega_2=2\pi \times 140~{\rm Hz}$. 
	(b) The Rb-PM output phase variations $\delta\varphi_{\rm PM}(t)$ with demodulation phase $\theta_{\rm lock}(t)$. The red curve represents output phase variations $\delta\varphi_{\rm PM}(t)$ for a second LCP calibration signal with frequency $\omega_3 = 2\pi \times 330~{\rm Hz}$.
	(c) The relative change of the Rb spin relaxation rate $\delta \Gamma_{\rm Rb}(t)/\Gamma_{\rm Rb}$ calculated according to Eqs.~\eqref{Eq:ComparePhase1}.
	Each data point represents a $\tau_{\rm ave} = 100~{\rm s}$ moving average of the raw data.
	}
	\label{fig:PMCloseLoop}
\end{figure}

Now we turn to the Xe NMR phase measured by the Rb-PM. 
Because of the opposite signs of the gyromagnetic ratios (i.e., $\gamma_{129}< 0$ and $\gamma_{131}>0$), the polarity of the NMR precession signals of the two isotopes are opposite.
The NMR signal of $^{129}$Xe is RCP, while the NMR signal of $^{131}$Xe is LCP.
The Rb-PM phases of $^{129}$Xe and $^{131}$Xe are
\begin{eqnarray}
	\varphi_{129}(\theta_1; t) &=& -k_p(\eta) \left(\theta_1 - \Theta_p^{(y)}(t)\right) - \frac{\omega_{129}}{\Gamma_{\rm Rb}},\\
	\varphi_{131}(\theta_1; t) &=&  k_p(\eta) \left(\theta_1 - \Theta_p^{(y)}(t)\right) - \frac{\omega_{131}}{\Gamma_{\rm Rb}},
\end{eqnarray}
To eliminate the change of the Rb-PM phases induced by the WP drift, $\Theta_p^{(y)}(t) =\Theta_p^{(y)}(t_0) + \delta \Theta_p^{(y)}(t)$, an RCP calibration signal is applied, which yields a Rb-PM phase 
\begin{equation}
	\varphi_{\rm cal}(\theta_1; t) = -k_p(\eta) \left(\theta_1 - \Theta_p^{(y)}(t)\right) - \frac{\omega_{\rm cal}}{\Gamma_{\rm Rb}}.
\end{equation}
The RCP calibration signal monitors the Rb-PM phase drift. 
To stabilize the Rb-PM output phase, we lock the phase $\varphi_{\rm cal}(\theta_1)$ of the calibration signal to zero by adjusting the demodulation phase $\theta_1$ via a feedback control loop.
In this case, the demodulation phase $\theta_1 = \theta_{\rm lock}(t)$ compensates the drift of the Rb-PM in real time, so that
\begin{equation}
\varphi_{\rm cal}(\theta_{\rm lock}(t) ) = 0.
\end{equation}
With the locking demodulation phase $\theta_{\rm lock}(t)$, the calibrated Rb-PM phase of the $^{129}$Xe and $^{131}$Xe signals are

\begin{eqnarray}
	\varphi_{129}(\theta_{\rm lock}) &=& \frac{\omega_{\rm cal} - \omega_{129}}{\Gamma_{\rm Rb}},\\
	\varphi_{131}(\theta_{\rm lock}) &=& - \frac{\omega_{\rm cal} + \omega_{131}}{\Gamma_{\rm Rb}}.
\end{eqnarray}
The phases are insensitive to the change of parametric modulation phase $\theta_{\rm ac}$ and the sensitivity to the change of magnetic field $B_0$ is much weaker.
The phase is only affected by the change of Rb spin relaxation rate $\delta \Gamma_{\rm Rb}(t)$, which causes the colored phase noise discussed in Sect~\ref{Sect:PhaseNoiseSpectrum}
\begin{eqnarray}
	\delta \varphi^{(129)}_{\rm c}(t) &\approx& \frac{\omega_{129} - \omega_{\rm cal} }{\Gamma_{\rm Rb}} \frac{\delta \Gamma_{\rm Rb}(t)}{\Gamma_{\rm Rb}},
	\label{Eq:PhaseNoise129}\\
	\delta \varphi^{(131)}_{\rm c}(t) &\approx& \frac{\omega_{131} + \omega_{\rm cal} }{\Gamma_{\rm Rb}} \frac{\delta \Gamma_{\rm Rb}(t)}{\Gamma_{\rm Rb}}.
	\label{Eq:PhaseNoise131}
\end{eqnarray}
In Eqs.~\eqref{Eq:PhaseNoise129} and \eqref{Eq:PhaseNoise131}, we have neglected the noise $\gamma_{129/131}\delta B_0/\Gamma_{\rm Rb}$ induced by the change of magnetic field $B_0$,
since they are too small (typically $\lesssim 10^{-6}~{\rm rad}$) comparing with the noise induced by the change of Rb spin relaxation rate $\delta \Gamma_{\rm Rb}$.

Equations~\eqref{Eq:PhaseNoise129} and \eqref{Eq:PhaseNoise131} indicate that, with the self-calibrating Rb-PM, 
the colored phase noise of Xe nuclear spins arises from the relative change of the Rb spin relaxation, i.e.,
\begin{equation}
\delta \varphi_{\rm c} (t) = \delta \Gamma_{\rm Rb}(t)/\Gamma_{\rm Rb},
\end{equation}
and the amplitudes are
\begin{eqnarray}
	\lambda_{129} = \frac{\omega_{129} - \omega_{\rm cal}}{\Gamma_{\rm Rb}},\\
	\lambda_{131} = \frac{\omega_{131} + \omega_{\rm cal}}{\Gamma_{\rm Rb}}.
\end{eqnarray}
According to Eq.~\eqref{Eq:suppresionFactor}, the amplitude suppression factor for a given frequency $\omega_{\rm cal}$ is
\begin{equation}
	\xi_{\omega_{\rm cal}} = \frac{(\Gamma_{129} - \Gamma_{131})\omega_{129}-(\Gamma_{129} + R\Gamma_{131})\omega_{\rm cal}}{(1+R)\Gamma_{\rm Rb}}.
\end{equation}
The amplitude $\xi_{\omega_{\rm cal}}$ linearly scales with the frequency of the calibration signal $\omega_{\rm cal}$.
The self-calibrating method offers a degree of freedom to continuously tune the amplitude of the colored phase noise via $\xi_{\omega_{\rm cal}}$.
Indeed, one can choose a critical frequency
\begin{equation}\label{optical_cal}
	\omega_{\rm cal}^{*} = \frac{\Gamma_{129} - \Gamma_{131} }{\Gamma_{129} + R\Gamma_{131} }\omega_{129},
\end{equation}
such that the suppression factor vanishes ($\xi_{\omega_{\rm cal}^*} = 0$).
However, due to the colored noise induced by the differential polarization field $\delta b_A$, $\omega_{\rm cal}^{*}$ is usually not the optimal choice for achieving the overall gyroscope stability (see Sect.~\ref{Sect:NMRGStability} below).

To validate the feasibility of the self-calibrating approach, we introduce two LCP calibration fields with frequency $\omega_2=2\pi\times 140$ Hz and $\omega_3=2\pi\times 330$ Hz,  respectively.
The field with frequency $\omega_2$ is used for self-calibrating process, i.e. locking the Rb-PM phase output $\delta\varphi_\mathrm{PM}|_{\omega_2}$ to zero,  as illustrated in FIG.~\ref{fig:PMCloseLoop}~(a) and (b). 
With the demodulation phase $\theta_\mathrm{lock}(t)$, the drift of the Rb-PM output phase for field $\omega_3$ is greatly suppressed compared to that without self-calibrating process (see FIG.~\ref{fig:PMOpenLoop}). 
The phase drift has been reduced to approximately $5~{\rm mdeg}$, a significant improvement from the $100~{\rm mdeg}$ shown in FIG.~\ref{fig:PMOpenLoop}. 
The remaining colored phase noise is primarily attributed to the relative changes of the Rb spin relaxation rate $\delta\Gamma_\mathrm{Rb}/\Gamma_\mathrm{Rb}$. 

The demodulation phase $\theta_{\rm lock}(t)$ compensates for the slow drift of the WP, but at the cost of introducing white noise to the NMR phases.
Together with the colored phase noise,  the white phase noise of the calibration signal also propagates to the NMR channels.
While the colored noises cancel each other out, the uncorrelated nature of white noise results in their cumulative effect.
With the white phase noise spectrum of the NMR phases $S_{\varphi, {\rm w}}^{(\alpha)}$ and the calibration signal $S_{\varphi, {\rm w}}^{(\rm cal)}$, 
the spectrum of the gyroscope signal induced by the phase noise is 
\begin{eqnarray}
\label{Eq:whiteSpectrum}
S_\mathrm{gyro}^{(\varphi)}(f)&=&\frac{|P_{129}(2\pi i f)|^2}{(1+R)^2}(S_{\varphi,\mathrm{w}}^{(129)}+S_{\varphi,\mathrm{w}}^{(\mathrm{cal})})\\ \nonumber
&&+\frac{|P_{131}(2\pi i f)|^2}{(1+R)^2}(S_{\varphi,\mathrm{w}}^{(131)}+S_{\varphi,\mathrm{w}}^{(\mathrm{cal})}).
\end{eqnarray}
In the low frequency limit, the angle random walk (ARW) of the NMRG is \cite{804271}
\begin{equation}
	\label{Eq:AWG}
	{\rm ARW} = \left[\frac{\tilde{\Gamma}_{129}^2 S_{\varphi,\mathrm{w}}^{(129)} +  \tilde{\Gamma}_{131}^2 S_{\varphi,\mathrm{w}}^{(131)}  + (\tilde{\Gamma}_{129}^2+\tilde{\Gamma}_{131}^2) S_{\varphi,\mathrm{w}}^{(\rm cal)}}{2} \right]^{1/2},
\end{equation}
where $\tilde{\Gamma}_{129} =  \Gamma_{129}/(1+R)$, $\tilde{\Gamma}_{131} =  R\Gamma_{131}/(1+R)$.
In our NMRG system, the phase noise $S_{\varphi,\mathrm{w}}^{(129)}  = (1.4~{\rm mdeg})^2/{\rm Hz}$, 
$S_{\varphi,\mathrm{w}}^{(131)} = (6.1~{\rm mdeg})^2/{\rm Hz}$, and $S_{\varphi,\mathrm{w}}^{(\rm cal)} = (5.2~{\rm mdeg})^2/{\rm Hz}$.
In fact, the phase noise of the calibration signal can be greatly reduced by increasing the signal amplitude.
Thus, the $^{131}$Xe phase noise usually dominates the ARW of our NMRG system.


\section{NMRG Stability}
\label{Sect:NMRGStability}

According to Eqs.~\eqref{Eq:Gyro_vs_bA} \& \eqref{Eq:phase_noise}, the low-frequency noise of the NMRG originates from 
the frequency noise induced by the fluctuation $\delta b_A(t)$ of the differential polarization field and the Rb-PM phase noise induced by the change of the Rb spin relaxation rate $\delta \Gamma_{\rm Rb}(t)/\Gamma_{\rm Rb}$, i.e.,
\begin{equation}
	\delta\tilde{\Omega}_{\rm gyro}(t) = \bar{\gamma}_{\rm Xe} \delta b_A(t) + \xi_{\omega_{\rm cal}}\frac{\delta \Gamma_{\rm Rb}(t)}{\Gamma_{\rm Rb}}.
\end{equation}
The changes of $\delta b_A(t)$ and $\delta \Gamma_{\rm Rb}(t)$ are the physical consequence of the drift of control parameters, e.g., the cell temperature $T$, the power $P$ the pump beam, etc..
Thus, the low-frequency drift of NMRG is expressed in terms of the changes of these parameters as
\begin{equation}
	\delta\tilde{\Omega}_{\rm gyro}(t) =\sum_{X} \left( \bar{\gamma}_{\rm Xe}\mathcal{K}_X + \xi_{\omega_{\rm cal}} \mathcal{Q}_X\right)\delta X(t),
	\label{Eq:GyroSig_vs_X}
\end{equation}
where 
\begin{eqnarray}
	\mathcal{K}_X &=& \frac{\partial b_A}{\partial X},\\
	\mathcal{Q}_X &=& \frac{\partial }{\partial X}\left( \frac{\delta \Gamma_{\rm Rb}}{\Gamma_{\rm Rb}}\right),
\end{eqnarray}
and $X$ stands for all possible parameters which affect $b_A$ and $\Gamma_{\rm Rb}$, i.e., $X\in \{T, P, \dots \}$.

\begin{figure*}[tbh]
	\centering
	\includegraphics[scale=1]{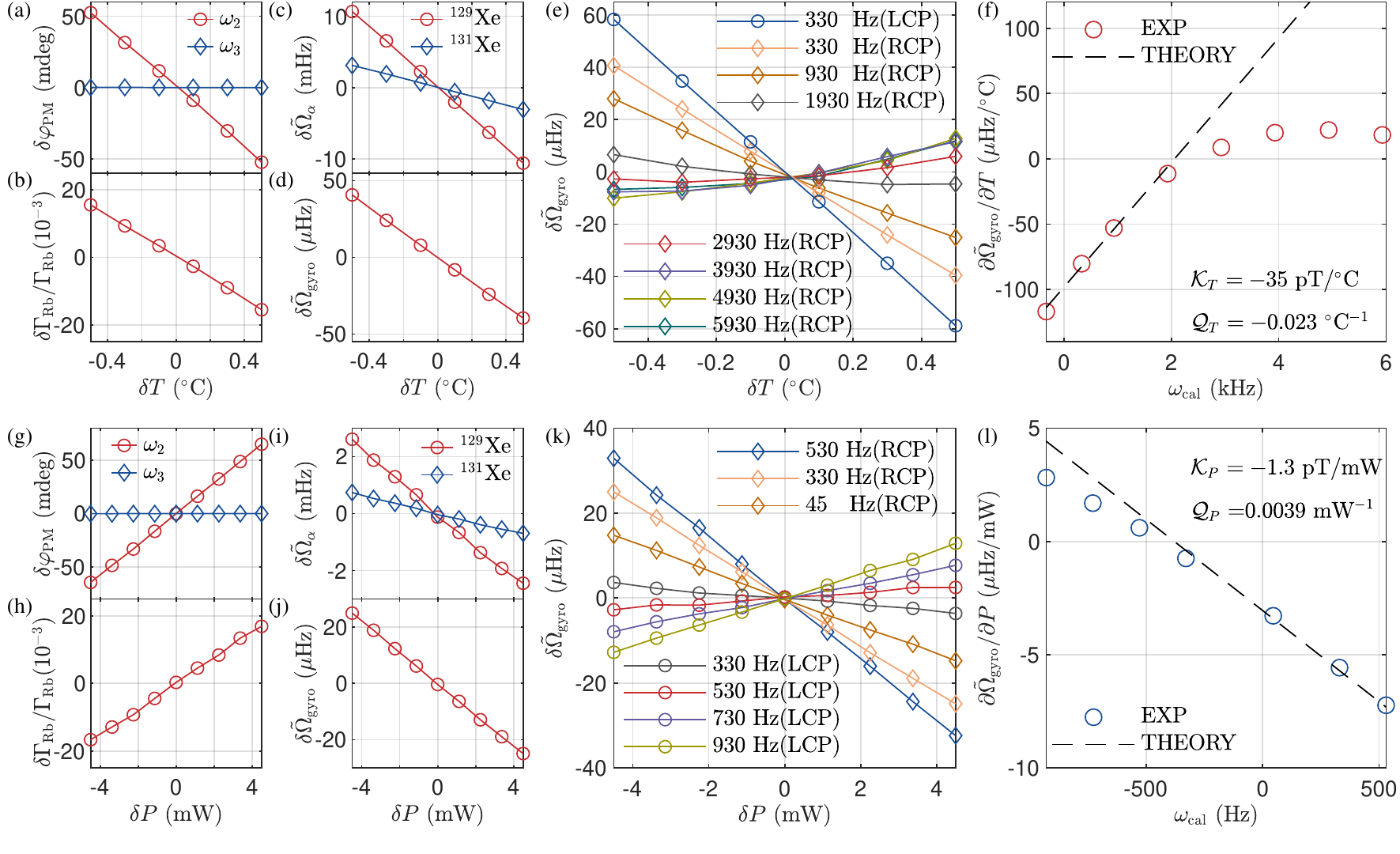}
	\caption{
	(a) \& (b) The Rb-PM phase output and the relative change of Rb spin relaxation rate $\delta \Gamma_{\rm Rb}/\Gamma_{\rm Rb}$ while sweeping the cell temperature. 
	A RCP calibration signal  with frequency $\omega_3=2\pi\times 330$ Hz is used for the self-calibrating. 
	Together with another LCP signal with frequency $\omega_2=2\pi\times 140$ Hz, $\delta \Gamma_{\rm Rb}/\Gamma_{\rm Rb}$ is derived according to Eqs.~\eqref{Eq:ComparePhase2}. 
	The slope in (b) gives $\mathcal{Q}_T=\partial(\delta \Gamma_{\rm Rb}/\Gamma_{\rm Rb})/{\partial T}$ in Eqs.~\eqref{Eq:GyroSig_vs_X}.
	(c) \& (d) The isotope NMR frequency shift $\delta \tilde{\Omega}_\alpha$ and the gyroscope signal drift $\delta \tilde{\Omega}_\mathrm{gyro}$ while sweeping the cell temperature. 
	The slope is $\delta \tilde{\Omega}_\mathrm{gyro}/{\delta T}=\bar{\gamma}_\mathrm{Xe}\mathcal{K}_\mathrm{T}+\xi_\mathrm{cal}\mathcal{Q}_\mathrm{T}$ with $\omega_\mathrm{cal}=\omega_3$.
	(e) The cell temperature dependence of the gyroscope signal with various frequencies of the calibration signal.
	(f) The slope of the gyroscope signal with respect to the cell temperature as a function of the frequency of the calibration signal.
	The symbols are extracted from the measured data in (e), and the dashed line is the theoretical prediction [according to Eq.~\eqref{Eq:GyroSig_vs_X}, with the control parameter $X=T$ and the slope determined in (b)].
	The parameter $\mathcal{K}_T$ is obtained from the intercept of the dashed line.
	(g) - (l) The same as (a) - (f), but for the control parameter $X=P$.
	The deviation of the measured data from the theoretical line in the region $\omega_{\rm cal}/(2\pi) > 2~{\rm kHz}$ is due to the violation of the condition $\omega_{\rm cal }/\Gamma_{\rm Rb} \ll 1$. 
	Other relevant parameters involved in these figures are: $\Gamma_{129}=2\pi\times 26.7$ mHz, $\Gamma_{131}=2\pi\times 12.6$ mHz, $\omega_{129}=2\pi\times 220$ Hz, $\omega_{131}=2\pi\times 65$ Hz, and $\Gamma_\mathrm{Rb}=2\pi\times 7.5$ kHz.
	}
	\label{fig:GyroscopeCalibration}
\end{figure*}

The self-calibrating Rb-PM provides a tool for analysing the NMRG drift.
Since $\xi_{\omega_{\rm cal}}$ is a tunable factor, for a given parameter $X$, we can control its contribution to the NMRG drift by choosing the frequency $\omega_{\rm cal}$ of the calibration signal.
Figure~\ref{fig:GyroscopeCalibration} shows the change of the NMRG output signal $\delta \tilde{\Omega}_{\rm gyro}$ as functions of the change of cell temperature $\delta T$ and power of the pump beam $\delta P$, with different calibration frequencies $\omega_{\rm cal}$.
The slope $\delta \tilde{\Omega}_{\rm gyro}/\delta X$ linearly scales with the calibration frequency $\omega_{\rm cal}$, from which we extract the parameters $\mathcal{Q}_X$ and $\mathcal{K}_X$.
Alternatively, the parameter $\mathcal{Q}_X$ can also be measured directly by comparing the phase change of multiple calibration signals while changing the parameter X [see Eqs.~\eqref{Eq:ComparePhase1} \& \eqref{Eq:ComparePhase2}].
The values of $\mathcal{Q}_X$ obtained by the two methods agree well.

\begin{table}
	\centering
	\caption{Parameters used in Eq.~\eqref{Eq:BI}}
	\begin{tabular}{p{0.08\textwidth}p{0.08\textwidth}p{0.08\textwidth}p{0.18\textwidth}}
	\hline \hline
	Quantity & Value & Unit  & Remark\\ 
	\hline
	$\mathcal{K}_T$              & -35 & $\rm pT/{}^{\circ}C$ & measured from Fig.~\ref{fig:GyroscopeCalibration}\\
	$\mathcal{K}_P$ & -1.3 & $\rm pT/mW$ & measured from Fig.~\ref{fig:GyroscopeCalibration}\\
	$\mathcal{Q}_T$              & -23 & $\rm \permil/{}^{\circ}C$ & measured from Fig.~\ref{fig:GyroscopeCalibration}\\
	$\mathcal{Q}_P$ & 3.9 & $\rm \permil/mW$ & measured from Fig.~\ref{fig:GyroscopeCalibration}\\
	$\bar{\gamma}_{\rm Xe}$      & 2.712 & $\rm {\mu Hz/pT}$ & defined in Eq.~\eqref{Eq:bar_gamma_xe}\\
	$h_T$ & $0.0015$ & ${\rm {}^{\circ}C}$ & typical value\\
	$h_{P}$ & $0.015$ & ${\rm mW}$ & typical value\\
	\hline \hline
	\end{tabular}
	\label{Table:Parameters}
\end{table}

\begin{figure}[tbh]
	\centering
	\includegraphics[scale=1]{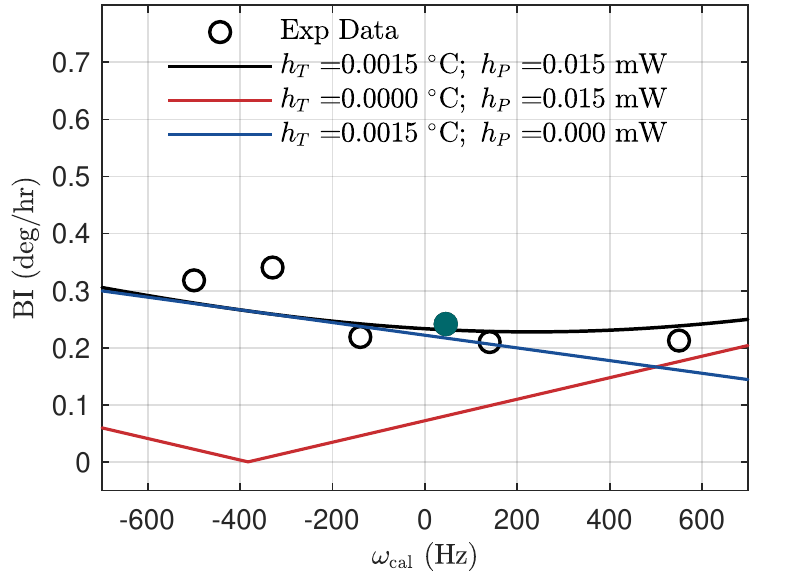}
	\caption{NMRG bias instability as a function of calibration frequency.
	Symbols are measured data and curves are theoretical results with different drift strengths $h_T$ and $h_P$. 
	The solid circle represents the BI value corresponding to the theoretically optimal calibration frequency $\omega_\mathrm{cal}^{*}$.}
	\label{fig:BI_vs_omegaCal}
\end{figure}

With the low-frequency spectrum of the parameter $\delta X(t)$ of the $1/f$ form $S_X(f) = h_X^2/f$,
and further assuming that the drift of different parameters are uncorrelated,
we obtain the low-frequency spectrum of NMRG output signal 
\begin{equation}
	S_{\rm gyro}(f) = \frac{\sum_{X} \left( \bar{\gamma}_{\rm Xe}\mathcal{K}_X + \xi_{\omega_{\rm cal}} \mathcal{Q}_X\right)^2 h_X^2}{f}.
\end{equation}
With the $1/f$ spectrum strength, the NMRG bias instability (BI) expressed in terms of the minimal Allan deviation in the $\sigma$-$\tau$ plot is \cite{804271}
\begin{equation}
	{\rm BI} = \left[2\ln 2 \sum_{X} \left( \bar{\gamma}_{\rm Xe}\mathcal{K}_X + \xi_{\omega_{\rm cal}} \mathcal{Q}_X\right)^2 h_X^2\right]^{1/2}.
	\label{Eq:BI}
\end{equation}
Assuming the cell temperature $T$ and the power $P$ of the pump beam are the dominating parameters causing the instability, i.e., $X\in\{T, P\}$,
with the parameters listed in Table~\ref{Table:Parameters}, 
we find the BI is insensitive to the calibration frequency within the range $\vert \omega_{\rm cal}\vert \le 2\pi \times 1~{\rm kHz}$ according to Eq.~\eqref{Eq:BI}, 
which agrees with the measured data in Fig.~\ref{fig:BI_vs_omegaCal}. 

\begin{figure*}[tbh]
	\centering
	\includegraphics[scale=1]{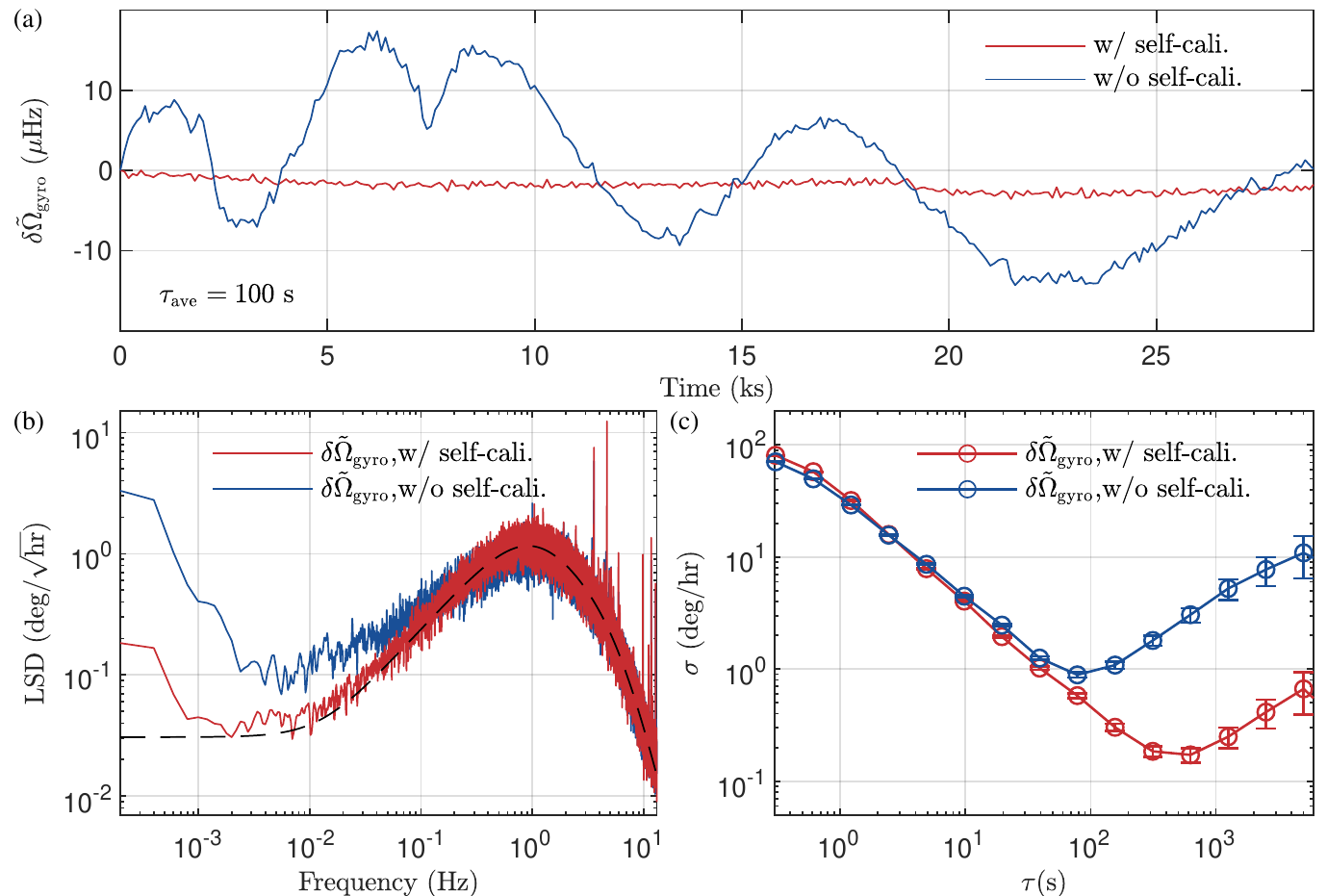}
	\caption{
	(a) The gyroscope signal with and without applying the self-calibrating Rb-PM. 
	A LCP calibration signal with frequency $\omega_\mathrm{cal}=2\pi\times 140$ Hz is applied and the gyroscope signal is continuously monitored for a duration of 8 hours. 
	The red curve depicts the gyroscope signal drift with a self-calibrating process, whereas no self-calibrating method is employed for the blue curve. 
	Each data point represents a $\tau_{\rm ave} = 100~{\rm s}$ moving average of the raw data.
	(b) The noise spectrum of the gyroscope signal with and without the self-calibrating Rb-PM. 
	The dashed curve is the noise spectrum due to the white phase noise, calculated according to Eq.~\eqref{Eq:whiteSpectrum}.
	(c) The Allan deviation of the gyroscope signal with and without the self-calibrating Rb-PM.
	}
	\label{fig:GyroscopeStability}
\end{figure*}

The NMRG with self-calibrating Rb-PM is experimentally implemented, as demonstrated in FIG.~\ref{fig:GyroscopeStability}. 
By using self-calibrating method, the gyroscope signal is stabilized within $\sim 3\mu$Hz for a duration of $8$ hours, 
one order of magnitude better than the case without the self-calibrating method. 
Figure~\ref{fig:GyroscopeStability} (b) compares the gyroscope performance with and without the self-calibrating method in the frequency domain. 
The colored noise in the low frequency range ($f\lesssim 1~{\rm mHz}$) is significantly reduced.
Figure~\ref{fig:GyroscopeStability} (c) shows the $\sigma$-$\tau$ plot of the Allan deviation.
With the self-calibrating method, the ${\rm BI} = 0.2^{\circ}/{\rm hr}$ is achieved in our system.

\section{Conclusion and outlook}
\label{Sect:Conclusion}

Improving system stability is one of the most important and challenging tasks in developing the NMRG.
The main difficulty is the lack of a precise model which describes the influence of various control parameters on the NMRG signal.
In this paper, we present a comprehensive analysis of the NMRG noise, including the white noise and colored noise.
Particularly, for the colored noise, two types of low-frequency noise source are discussed, namely, the frequency noise $\delta b_A$ induced by the differential polarization field 
and the phase noise $\delta \varphi_{\rm PM}$ induced by the Rb-PM.

We develop the exact solution to the equation of motion of the Rb-PM. 
With the exact solution, the behaviour of the Rb-PM phase $\varphi_{\rm PM}$ can be understood with a high precision down to $\sim 10^{-3}{\rm deg}$.
The analytic results are confirmed by our experiment measurements.
Also, based on the solution, we propose and implement the self-calibrating method to compensate one of the main low-frequency phase noise, i.e., the drift of the WP.
The self-calibrating Rb-PM significantly improves the NMRG stability.

The self-calibrating Rb-PM also provides a tool for diagnosing the NMRG system.
The amplitude of the Rb-PM phase noise can be tuned by choosing the frequency of the calibration signal. 
With this degree of freedom, hidden information, like the temperature and pump power dependence of the differential polarization field, is extracted from the measured data.
The knowledge about the differential polarization field is valuable for the further improvement of the NMRG stability.

Due to the limited space of this paper, the physical origin of the differential polarization field and its dependence on the control parameters will be discussed in another work.
With the deep understanding of the differential polarization field, and the experimental tool developed in this work, the NMRG performance would be hopefully further improved.

\begin{acknowledgments}
We thank Professor Dong Sheng for the inspiring discussion. We thank Dr. Bowen Song for her previous work on the experimental setup.
We thank Kang Dai for providing the vapor cell. This work is supported by NSFC (Grants No.~U2030209, No.~52007177, No.~12088101 and No.~U223040003).

\end{acknowledgments}

\appendix

\section{The stationary response of the Rb-PM}\label{Appendix:Rb-PM Response}
The dynamics of the transverse component $\langle S_+ \rangle$ is governed by
\begin{align}\label{Appendix_BlochEq}
  \frac{d}{dt}\langle S_+\rangle=&\left[-i\Omega_0-i \gamma_{\mathrm{Rb}}B_{\mathrm{ac}}\cos\left({\omega_0 t+\theta_{\mathrm{ac}}}\right)-\Gamma_{\mathrm{Rb}}\right]\langle S_+\rangle \nonumber \\
  +&i\gamma_{\mathrm{Rb}}b_c(t)\langle S_z\rangle
\end{align}
and the solution has been derived by other researchers with various approximations. In this paper, we go beyond the adiabatic approximation, which takes $b_c(t)$ as time independent, and give a more accurate solution to analyze the NMRG stability.

Since Eqs.~\eqref{Appendix_BlochEq} is a first-order linear differential equation and only stationary solutions are necessary in experimental analysis, we can consider the stationary solutions of Eqs. \eqref{Appendix_BlochEq} to the positive frequency and negative frequency component of $b_c(t)$ separately. The response of $\langle S_+\rangle$ to the positive frequency component $b^+ \exp\left({i \omega t}\right)$ follows the equation
\begin{align}\label{Appendix_BlochEqPlus}
  \frac{d}{dt}\langle S_+\rangle=&\left[-i\Omega_0-i \gamma_{\mathrm{Rb}}B_{\mathrm{ac}}\cos\left({\omega_0 t+\theta_{\mathrm{ac}}}\right)-\Gamma_{\mathrm{Rb}}\right]\langle S_+\rangle \nonumber \\
  +&i\gamma_{\mathrm{Rb}}b^{+}e^{i\omega t}\langle S_z\rangle
\end{align}
By defining
\begin{align}\label{Bloch_Trans}
\langle \tilde{S}_+\rangle=\langle S_+\rangle e^{i[\Omega_0 t+\eta \sin(\omega_0 t+\theta_{\mathrm{ac}})+\Gamma_{\mathrm{Rb}}t]}
\end{align}
and using the Jacobi-Anger expansion
\begin{align}
e^{iz \sin{\theta}}=\sum_{n=-\infty}^{\infty} J_n(z) e^{i\theta z},
\end{align}
we get the simplified differential equation
\begin{align}\label{Bloch_EqSimp}
\frac{d}{dt}\langle \tilde{S}_+ \rangle =i \gamma_{\mathrm{Rb}}b^+ \langle S_z \rangle \sum_{n=-\infty}^{\infty} J_n(\eta)e^{i(\Omega_0+n\omega_0+\omega)t+\Gamma_{\mathrm{Rb}}t+in\theta_{\mathrm{ac}}}.
\end{align}
Integrate both sides of Eqs.~\eqref{Bloch_EqSimp} from $(0,t)$ and use Eqs.~\eqref{Bloch_Trans}, we get the stationary response of the Rb-PM to the positive frequency field
\begin{align}
\langle S_+\rangle_{+\omega}=\frac{\gamma_{\mathrm{Rb}}}{\Gamma_{\mathrm{Rb}}}\langle S_z \rangle b^{+} e^{+i \omega t}\sum_{p=-\infty}^{\infty} \mathcal{A}_p^+(\eta,\Omega_0,\omega_0,\omega)e^{ip(\omega_0 t+\theta_{\mathrm{ac}})}.
\end{align}
The stationary response of the Rb-PM to the negative frequency field can be derived in the same manner with
\begin{align}
\langle S_+\rangle_{-\omega}=\frac{\gamma_{\mathrm{Rb}}}{\Gamma_{\mathrm{Rb}}}\langle S_z \rangle b^{-} e^{-i \omega t}\sum_{p=-\infty}^{\infty} \mathcal{A}_p^{-}(\eta,\Omega_0,\omega_0,\omega)e^{ip(\omega_0 t+\theta_{\mathrm{ac}})}.
\end{align}
The total stationary response of the Rb-PM to the magnetic field $b_c(t)=b^+ \exp\left({i \omega t}\right)+b^- \exp\left({-i\omega t}\right)$ is the summation of the positive frequency response $\langle S_+\rangle_{+\omega}$ and negative frequency response $\langle S_+\rangle_{-\omega}$, i.e.
\begin{align}
\langle S_+ \rangle = \frac{\gamma_\mathrm{Rb}}{\Gamma_{\mathrm{Rb}}} \langle S_z \rangle \sum_{p=-\infty}^{\infty} \left(b^+ e^{i\omega t} \mathcal{A}_p^+ +b^- e^{-i\omega t}\mathcal{A}_p^-\right)e^{ip(\omega_0 t+\theta_\mathrm{ac})},
\end{align}
which gives
\begin{align}
\langle S_x \rangle=\frac{\gamma_\mathrm{Rb}\langle S_z \rangle}{2\Gamma_{\mathrm{Rb}}} \sum_{p=-\infty}^{\infty} \left(b^+ e^{i\omega t} \mathcal{A}_p^+ +b^- e^{-i\omega t}\mathcal{A}_p^-\right)e^{ip(\omega_0 t+\theta_\mathrm{ac})}+{\rm c.c.} .
\end{align}

\section{The Gain functions}
\label{Appendix:Gainfunctions}

The gain functions $G_p^{(x/y/\pm)}(\theta_1)$ are crucial in analysing the NMRG stability. 
We present a detailed discussion on the gain functions.
With the dimensionless parameters $\zeta = \omega_0/\Gamma_{\rm Rb}$, $\delta = \Delta_{\rm Rb}/\Gamma_{\rm Rb}$ and $x = \omega/\Gamma_{\rm Rb}$,
the complex amplitudes $\mathcal{A}_p^{\pm}$ are expressed as
\begin{equation}
	\mathcal{A}_p^{\pm}=\frac{  J_{-p-1}(\eta) J_{-1}(\eta) }{\delta \pm x-i} + \sum_{n\neq 0} \frac{  J_{n-p-1}(\eta) J_{n-1}(\eta) }{ n\zeta + \delta \pm x-i}.
\end{equation}
By expanding $\mathcal{A}_p^{\pm}$ to the second order of $\zeta^{-1}$, $\delta$, and $x$, we get
\begin{eqnarray}
	&&G^{(x)}_{p}(\theta_1) =-\left[\left(1-\delta^2-x^2\right)d_p+\zeta^{-2}d_p''\right]\sin\theta_p\\
 				&&+i\left(-2s_p\delta x\cos\theta_p+d_px\sin\theta_p\right)+\left(s_p\delta+s_p'\zeta^{-1}\right)\cos\theta_p,\notag\\ 
	&&G^{(y)}_{p}(\theta_1) = -i(2d_p\delta x\sin\theta_p+s_px\cos\theta_p)\\
				&&+\left(d_p\delta + d_p'\zeta^{-1}\right)\sin\theta_p+\left[\left(1-\delta^2-x^2\right)s_p+\zeta^{-2}s_p''\right]\cos\theta_p,\notag	
\end{eqnarray}
where we have defined $\theta_p=\theta_1-p\theta_{\rm ac}$, and
\begin{eqnarray}
	s''_p(\eta) &= 2\sum_{n\neq 0} n^{-2}J_{n-1}(\eta)\left[J_{n+p-1}(\eta)+J_{n-p-1}(\eta)\right],\\
	d''_p(\eta) &= 2\sum_{n\neq 0} n^{-2}J_{n-1}(\eta)\left[J_{n+p-1}(\eta)-J_{n-p-1}(\eta)\right].
\end{eqnarray}
Up to the second order corrections, the magnitudes of $G_p^{(x)}(\theta_1)$ and $G_p^{(y)}(\theta_1)$ are 
\begin{align}
\left|G_p^{x}\right|=&\left|d_p\right|\sqrt{ \sin^2(\theta_p-\phi_p^{(x)})+x^2(k_p\delta-k_p^{\prime}\zeta^{-1})^2}, \label{GpxNorm}\\
\left|G_p^{y}\right|=&\left|s_p\right|\sqrt{ \cos^2(\theta_p-\phi_p^{(x)})+x^2(q_p\delta-q_p^{\prime}\zeta^{-1})^2},\label{GpyNorm}
\end{align}
where the phases $\phi_p^{(x)}$ and $\phi_p^{(y)}$ are defined as
\begin{align}
\phi_p^{(x)}=& k_p \delta+k_p^{\prime}\zeta^{-1},\\
\phi_p^{(y)}=& q_p\delta+q_p^{\prime}\zeta^{-1}.
\end{align}
The response of the Rb-PM to the transverse fields can be tuned by choosing different demodulation phase $\theta_1$. 
Particularly, the response to $B_x$ and $B_y$ is minimized when $\theta_1$ takes the value 
\begin{align}
\theta_1=p\theta_{\mathrm{ac}}+k_p\delta+k_p^{\prime}\zeta^{-1}\equiv \Theta_p^{(x)},
\end{align}
or
\begin{align}
\theta_1=p\theta_{\mathrm{ac}}+q_p\delta+q_p^{\prime}\zeta^{-1}+\frac{\pi}{2}\equiv \Theta_p^{(y)}.
\end{align}

\begin{figure}
  \centering
  \includegraphics[width=8cm]{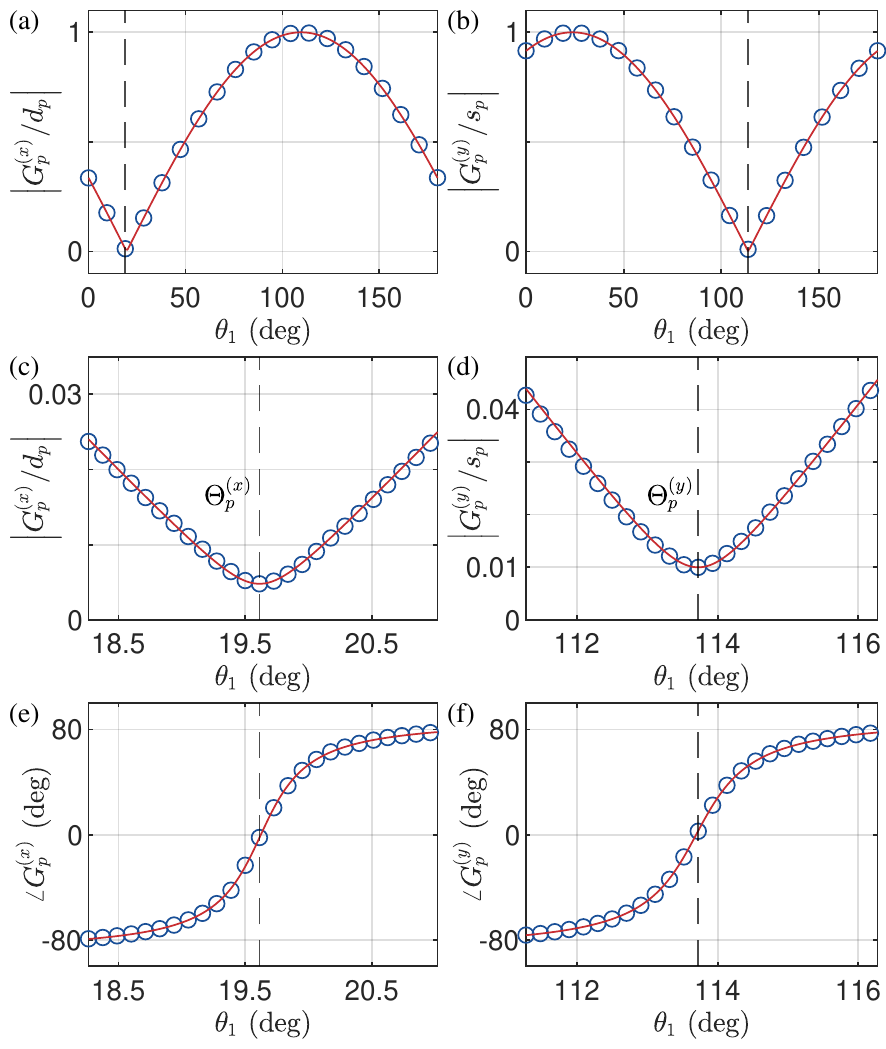}\\
  \caption{
	(a) \& (b) The normalized magnitude $G_p^{(x)}(\theta_1)$ and $G_p^{(y)}(\theta_1)$ as functions of the demodulation phase $\theta_1$. The symbols are experimental data, and the red curves are the theoretical results according to Eqs.~\eqref{GpxNorm} and \eqref{GpyNorm}.
	(c) \& (d) The magnified behaviour of $\left\vert G_p^{(x)}(\theta_1)\right\vert $ and $\left\vert G_p^{(y)}(\theta_1)\right\vert$ near $\Theta_p^{(x/y)}$.
        (e) \& (f) The phase $\arg[G_p^{(x)}(\theta_1)]$ and $\arg[G_p^{(y)}(\theta_1)]$ near $\Theta_p^{(x/y)}$. 
	The red curves are the theoretical results according to Eqs.~\eqref{Eq:GainFunctionXPhase} \&~\eqref{Eq:GainFunctionYPhase}.
  }\label{WorkingPointFinding}
\end{figure}

The phase angle $\arg [G_p^{(x)}(\theta_{1})]$ and $\arg [G_p^{(y)}(\theta_{1})]$ characterize the phase response to the transverse $B_{x}$ and $B_{y}$ respectively. 
Near the critical points $\Theta_p^{(x/y)}$, the phase $\arg[G_p^{(x/y)}]$ are approximated to
\begin{eqnarray}
	\arg[G_p^{(x)} (\theta_{1})] = \arctan\left[\frac{x\left(k_p\delta-k_p^{\prime}\zeta^{-1}\right)}{\theta_{1}-\Theta_p^{(x)}}\right],
	\label{Eq:GainFunctionXPhase}\\
	\arg[G_p^{(y)} (\theta_{1})] = \arctan\left[\frac{x\left(q_p\delta-q_p^{\prime}\zeta^{-1}\right)}{\theta_{1}-\Theta_p^{(y)}}\right].
	\label{Eq:GainFunctionYPhase}
\end{eqnarray}
Figure~\ref{WorkingPointFinding} compares the experimentally measured gain functions and the theoretically calculated values, which shows excellent agreement.

Since the NMR signals are LCP or RCP fields, the more relevant gain functions are actually $G_p^{(\pm)}(\theta_1)$. 
Up to the first order corrections, the gain functions for the circularly polarized fields are
\begin{eqnarray}
	&&G_p^{(+)}(\theta_1) = i\left[s_p\cos\theta_p+\left(\delta+x \right)d_p\sin\theta_p+\zeta^{-1}d_p'\sin\theta_p\right]\notag\\
		&&	   - d_p\sin\theta_p+\left(\delta+x \right)s_p\cos\theta_p+\zeta^{-1}s_p'\cos\theta_p,
\end{eqnarray}
and
\begin{eqnarray}
	&&G_p^{(-)}(\theta_1) = -i\left[s_p\cos\theta_p+\left(\delta-x \right)d_p\sin\theta_p+\zeta^{-1}d_p'\sin\theta_p\right]\notag\\
		&&	   - d_p\sin\theta_p+\left(\delta-x \right)s_p\cos\theta_p+\zeta^{-1}s_p'\cos\theta_p.
\end{eqnarray}
The phases of the gain functions are 
\begin{eqnarray}
	&&\arg[G_p^{(+)}(\theta_1)]= \label{Eq:GainFunctionLCPPhase}\\
	&&\arctan\left[\frac{s_p\cos\theta_p+\left(\delta+x \right)d_p\sin\theta_p+\zeta^{-1}d_p'\sin\theta_p}{-d_p\sin\theta_p+\left(\delta+x \right)s_p\cos\theta_p+\zeta^{-1}s_p'\cos\theta_p }\right],\notag
\end{eqnarray}
and 
\begin{eqnarray}
	&&\arg[G_p^{(-)}(\theta_1)]= \label{Eq:GainFunctionRCPPhase}\\
	&&\arctan\left[-\frac{s_p\cos\theta_p+\left(\delta-x \right)d_p\sin\theta_p+\zeta^{-1}d_p'\sin\theta_p}{-d_p\sin\theta_p+\left(\delta-x \right)s_p\cos\theta_p+\zeta^{-1}s_p'\cos\theta_p }\right]\notag.
\end{eqnarray}
Near the WP $\theta_1 \approx \Theta_p^{(y)}$, the phases are simplify as
\begin{eqnarray}
	\arg[G_p^{(+)}(\theta_1)] &=& \arctan\left[\frac{k_p\cot\theta_p+\left(\delta+x \right)+\zeta^{-1}d_p'/d_p}{-1+\left(\delta+x \right)k_p\cot\theta_p+\zeta^{-1}k_p'\cot\theta_p }\right]\notag\\
	&\approx& k_p(\theta_1 - \Theta_p^{(y)}) - x,
	\label{Eq:GainFunctionLCPPhaseWP}
\end{eqnarray}
and
\begin{eqnarray}
	\arg[G_p^{(-)}(\theta_1)] &=& \arctan\left[-\frac{k_p\cot\theta_p+\left(\delta-x \right)+\zeta^{-1}d_p'/d_p}{-1+\left(\delta-x \right)k_p\cot\theta_p+\zeta^{-1}k_p'\cot\theta_p }\right]\notag\\
	&\approx& -k_p(\theta_1 - \Theta_p^{(y)}) - x,
	\label{Eq:GainFunctionRCPPhaseWP}
\end{eqnarray}
which are Eq.~\eqref{Eq:PhaseAngle_Gpm} of the main text.
Here, we have used the facts that
\begin{eqnarray}
	\cot\theta_p &=& -\tan\left[\left(\theta_1-\Theta_p^{(y)}\right)+q_p\delta+q_p'\zeta^{-1}\right]\notag\\
		     &\approx&-(\theta_1-\Theta_p^{(y)})-q_p\delta-q_p'\zeta^{-1},
\end{eqnarray}
and the functions $k_p(\eta)$, $k_p'(\eta)$, $d_p(\eta)$, $d_p'(\eta)$, $q_p(\eta)$, and $q_p'(\eta)$ are all on the order of unity in our experimental setup with $\eta\approx 1$.

\providecommand{\noopsort}[1]{}\providecommand{\singleletter}[1]{#1}%
\end{document}